\documentclass[fleqn,usenatbib]{mnras}

\usepackage{newtxtext,newtxmath}

\usepackage[T1]{fontenc}

\DeclareRobustCommand{\VAN}[3]{#2}
\let\VANthebibliography\thebibliography
\def\thebibliography{\DeclareRobustCommand{\VAN}[3]{##3}\VANthebibliography}

\usepackage{graphicx}	
\usepackage{amsmath}	

\usepackage{tikz}

\title[Secular evolution of dust disc sizes]{Modelling the secular evolution of proto-planetary disc dust sizes -- A comparison between the viscous and magnetic wind case} 

\author[F.~Zagaria et al.]{
Francesco Zagaria,$^{1}$\thanks{E-mail: fz258@cam.ac.uk}
Giovanni P. Rosotti,$^{2,3}$
Cathie J. Clarke,$^{1}$
and Beno\^{i}t Tabone$^{4}$
\\
$^{1}$Institute of Astronomy, University of Cambridge, Madingley Road, Cambridge CB3 0HA, UK\\
$^{2}$School of Physics and Astronomy, University of Leicester, Leicester LE1 7RH, UK\\
$^{3}$Leiden Observatory, Leiden University, PO Box 9513, 2300 RA Leiden, The Netherlands\\
$^{4}$Université Paris-Saclay, CNRS, Institut d’Astrophysique Spatiale, 91405 Orsay, France\\
}

\date{Accepted XXX. Received YYY; in original form ZZZ}

\pubyear{2021}

\begin{document}
\label{firstpage}
\pagerange{\pageref{firstpage}--\pageref{lastpage}}
\maketitle

\begin{abstract}
    For many years proto-planetary discs have been thought to evolve viscously: angular momentum redistribution leads to accretion and outward disc spreading. Recently, the hypothesis that accretion is due, instead, to angular momentum removal by magnetic winds gained new popularity: no disc spreading is expected in this case. In this paper, we run several one-dimensional gas and dust simulations to make predictions on the time-evolution of disc sizes \textit{in the dust} and to assess whether they can be used to understand how discs evolve. We show that viscous and magnetic wind models have very different dust disc radii. In particular, MHD wind models are compact and their sizes either remain constant or decrease with time. On the contrary, discs become larger with time in the viscous case (when $\alpha\gtrsim10^{-3}$). Although current observations lack enough sensitivity to discriminate between these two scenarios, higher-sensitivity surveys could be fruitful to this goal on a $1\,{\rm to}\,10\,{\rm Myr}$ age range. When compared with the available ALMA Band~7 data, both viscous and magnetic wind models are compatible with the observationally-inferred dust radii in Lupus, Chamaeleon~I and Upper Sco. Furthermore, in the drift-dominated regime, the size-luminosity correlation is reproduced in Lupus, both in Band~7 and 3, while in Upper Sco a different slope than in the data is predicted. Sub-structures (potentially undetected) can explain several outliers with large observed sizes. Higher-angular-resolution observations will be helpful to test our predictions in the case of more compact discs, expected in both frameworks, particularly at the age of Upper Sco. 
\end{abstract}




\section{Introduction}\label{sec:1}
Planets form in discs of gas and dust orbiting young stars. Knowledge of the evolutionary processes that such planet-forming discs undergo before dissipating is essential to any planet-formation theory \citep[e.g.,][]{Morbidelli&Raymond2016} and key to understanding the properties of the currently observed exoplanets, such as their system architectures \citep{Winn&Fabricky2015}, their composition \citep{Madhusudhan2019}, and ultimately their potential to host life (e.g., by the presence of water and organics, \citealt{Oberg&Bergin2021}). 

Disc evolution is a long-standing problem motivated by the evidence that most new-born stars accrete gas \citep{Hartmann2016}. Since the pioneering work of \citet{Shakura&Sunyaev1973} and \citet{Lynden-Bell&Pringle1974}, proto-planetary discs have been thought to evolve under the effect of viscosity (an averaged effective turbulence). In this picture, viscosity allows for accretion onto the forming star transporting some material outwards, in a process known as \textit{viscous spreading}. Nevertheless, the physical origin and magnitude of viscosity are still debated (e.g., \citealt{Turner2014}). For this reason, \citet{Shakura&Sunyaev1973} introduced a parametrisation of viscosity in terms of a coefficient, $\alpha_{\rm SS}$, embodying our ignorance of the physical mechanisms behind turbulence. Traditionally, the magneto-rotational instability (MRI) has been invoked as a possible source of turbulence in the inner disc \citep{Balbus&Hawley1991}, while gravitational instability (GI) has been proposed for the massive outer disc regions, especially in the first phases of disc formation \citep{Kratter&Lodato2016}. 

However, in the last years it was shown that non-ideal magneto-hydrodynamical (MHD) effects, such as Ohmic and ambipolar diffusion, and the Hall drift, may lead to the quenching of MRI (e.g., the reviews of \citealt{Turner2014} and \citealt{Lesur2020}). Firstly, \citet{Gammie1996} showed that Ohmic resistivity can efficiently suppress MRI in the inner disc mid-plane (dead zone), and proposed that accretion could take place through the MRI-active upper disc layers. Later, \citet{Bai2013} and \citet{Bai&Stone2013} performed local shearing-box simulations of stratified discs where both Ohmic resistivity (dominating in the mid-plane) and ambipolar diffusion (dominating at the surface) were considered, finding that MRI is also quenched in the upper layers of the disc. As a solution to the angular momentum transport problem, these authors revived the old scenario of MHD disc winds, showing that in the presence of a magnetic field with net magnetic flux, accretion is efficiently driven by the launching of a magneto-centifugal wind. This evidence brought back to popularity the old scenario, firstly proposed by \citet{Blandford&Payne1982}, \citet{Wardle&Koenigl1993} and revised by \citet{Ferreira1997}, that accretion is due to MHD disc winds.

State-of-the-art global simulations, including a realistic treatment of sub-grid micro-physics, confirm these scenario, showing that, even when all the non-ideal MHD terms are considered, accretion can be efficiently driven by magnetic disc winds \citep[e.g.,][]{Bethune2017,Bai2017,Wang2019,Gressel2020}. What is more, recent observational results of different kind are also supporting the importance of MHD winds for disc evolution. On the one hand, we have evidence that turbulence is low in the outer disc regions (e.g., \citealt{Flaherty2017,Flaherty2018}, but see \citealt{Flaherty2020} for an exception). On the other hand, ALMA observations are also detecting MHD disc wind candidates in molecular lines (e.g., \citealt{deValon2020,Louvet2016,Tabone2017,Booth2021}) and forbidden atomic lines (e.g., \citealt{Fang2018,Banzatti2019,Pascucci2020,Whelan2021}, even though distinguishing between thermal and magnetic winds is not straightforward).

When disc evolution is dominated by MHD winds, \textit{no spreading} is expected because angular momentum is not radially redistributed but removed vertically. This suggests that confronting global disc sizes at different epochs can be a fruitful method to assess if viscosity or winds are the drivers of disc evolution. However, global non-ideal MHD simulations are difficult and numerically expensive to perform because both the large scale magnetic field and the sub-grid chemistry (to compute the dissipation coefficients) need to be modelled with care and be numerically resolved. Using the results of local simulations, global one-dimensional models can be built \citep[e.g.,][]{Armitage2013,Bai2016,Lesur2021}. Nevertheless, all such models are based on several assumptions owing to their dependence on the simulations to which
they were tailored.

Recently, \citet{Tabone2022a} introduced a new simple parametrisation of the problem. Following the idea of \citet{Suzuki2016} of applying the parametrisation of \citet{Shakura&Sunyaev1973} to magnetic winds, they introduced a coefficient, $\alpha_{\rm DW}$, embodying our ignorance on the magnitude of angular momentum removal by a MHD wind. In addition to \citet{Suzuki2016}, \citet{Tabone2022a} were able to find analytical self-similar solutions for wind-dominated discs, analogous to the viscous ones of \citet{Lynden-Bell&Pringle1974} in the viscous case. This makes a large-scale comparison between the two model predictions feasible. In particular, they confirmed the result of \citet{Armitage2013} that in the absence of viscosity no disc spreading takes place.

Since the advent of the Atacama Large Millimeter/submillimeter Array (ALMA), more and more nearby star-forming regions have been surveyed at low to moderate angular resolution \citep[e.g.,][]{Ansdell2016,Ansdell2018,Barenfeld2016,Pascucci2016}, making it possible to compute disc gas and dust sizes for a large number of sources \citep[e.g.,][]{Barenfeld2017,Tazzari2017,Ansdell2018,Sanchis2021}. This allows us to compare the different theoretical predictions for disc size evolution with time to data in star-forming regions of different ages in a statistical sense.

\citet{Najita&Bergin2018} were the first to use disc \textit{gas} sizes to search for evidence of viscous spreading between Class~0/I and Class~II discs, finding tentative evidence of more evolved discs being larger. However, their sample was non-homogeneous and different molecular tracers (CN and CO, mainly) were used to observationally infer gas sizes. Most importantly, disc sizes measured from CO (sub-)millimetre rotational emission cannot be naively compared to the predictions of viscous evolution, because they are affected by changing physical and chemical conditions over secular time scales. To address this problem, \citet{Trapman2020} built a set of thermo-chemical viscous models taking into account processes such as CO photo-dissociation, chemical processing (in the warm molecular layer), freeze-out on grains and radial transport, affecting the abundance, distribution and radial extension of CO emission. Their results showed that, when $\alpha_{\rm SS}=10^{-3}\,{\rm and}\,10^{-4}$ models and data agree for the young Lupus star-forming region, while for the older Upper Sco OB association, CO sizes were too small for their models (even though no sensitivity cut was considered in the post-processing). \citet{Trapman2022} performed a similar exercise using the magnetic wind models of \citet{Tabone2022a}, showing that a good agreement between numerical predictions and data can be seen both in Lupus and Upper Sco. However, magnetic wind models fail at reproducing the disc sizes of younger Class 0/I sources. 

In this paper we follow an alternative approach, and discuss whether disc \textit{dust} sizes can be used as a proxy for disc evolution. This choice can be motivated both by theoretical and observational arguments. In fact, solids dominate ALMA continuum observations, determining the opacity and temperature structure of the disc, and are the key ingredient for planet formation \citep{Birnstiel2016}. Moreover, at the time of writing, gas sizes have been measured only in 35 discs in Lupus \citep{Sanchis2021} and 9 in Upper Sco \citep{Barenfeld2017}, while for dust a larger, homogeneously analysed sample of 30 discs from Lupus, 33 from Chamaeleon~I, and 22 from Upper Sco is available \citep{Hendler2020}.

Relying on dust size measurements in the Lupus sample of \citet{Ansdell2016}, a tentative evidence of viscous spreading was reported by \citet{Tazzari2017}, who showed that Lupus discs were larger than those in the younger Taurus-Auriga and $\rho$~Ophiuchus star-forming regions. However, this trend was not confirmed by \citet{Andrews2018}. When dust is taken into account, several processes come into play that need to be addressed by detailed modelling. In particular, as highlighted by several authors \citep[e.g.,][]{Weidenchilling1977,Takeuchi&Lin2002,Birnstiel2010,Birnstiel&Andrews2014,Testi2014} the dynamics of solids is different from that of the gas because dust is subject to radial drift. Briefly, as a result of the damping effect of the gas, solid particles drift towards the star, with a speed proportional to their size, potentially shrinking the disc dust emission region. 

\citet{Rosotti2019_radii} showed that this is not the case in viscously evolving discs because, promoting the removal of large grains, radial drift becomes ``a victim of its own success''. In fact, once the larger solids are accreted, smaller particles, well coupled with the spreading gas and drifting outwards, determine the overall expansion of the dust disc size with time (when $\alpha_{\rm SS}\gtrsim 10^{-3}$). However, \citet{Rosotti2019_radii} showed that this trend of increasing dust sizes cannot be seen at the sensitivity of the current population surveys and would be challenging to detect even for much deeper observations.

In this paper we run one-dimensional gas and dust models, relying on the parametrisation of \citet{Tabone2022a} and the simplified treatment of dust growth of \citet{Birnstiel2012}. Following \citet{Rosotti2019_radii,Rosotti2019_fr}, our aim is to make predictions for the time-evolution of the dust disc sizes, assuming that disc evolution is ruled by either viscosity or magnetic winds. We then compare our results with the dust disc sizes inferred from the currently available ALMA observations to determine their agreement and whether this depends on the underlying disc evolution mechanism. We remark that previous works already took into account dust evolution in one-dimensional MHD disc wind models (e.g., \citealt{Takahashi&Muto2018,Taki2021,Arakawa2021}, using the prescriptions of \citealt{Suzuki2016} for gas evolution). However, to our knowledge, none of these works focused on disc sizes.

This paper is organised as follows. In Section~\ref{sec:2} we introduce our model. Section~\ref{sec:3} illustrates our results in a test case. The effects of magnetic winds on dust evolution are discussed, and how they impact the evolution of the mass and flux radius. Section~\ref{sec:4} deals with the parameter space exploration, while Section~\ref{sec:5} is dedicated to the observational consequences of our work: are current observations able to distinguish between viscous and magnetic-wind evolution, and how our predictions on the correlation between (sub-)millimetre disc sizes and fluxes change with magnetic winds? In Section~\ref{sec:6} we discuss possible model limitations. Finally, in Section~\ref{sec:7} we summarise our results and draw our conclusions.

\section{Numerical methods and observations}\label{sec:2}
One-dimensional simulations of gas and dust evolution are performed using the code developed by \citet{Booth2017}. The magnetic wind models of \citet{Tabone2022a} are used for gas, while for dust we employ the simplified treatment of grain growth developed by \citet{Birnstiel2012}. The outputs of our simulations are then post-processed to produce synthetic (sub-)millimetre dust observations, to be compared with real data. Hereafter, our numerical methods and observational sample are described.

\subsection{Numerical methods}
\paragraph*{Code architecture} The code implements by default a routine solving the viscous evolution equation \citep{Lynden-Bell&Pringle1974}. This is updated to take into account the effect of magnetic winds as in the master equation of \citet[][see Eq.~10, which is written here in the Keplerian case]{Tabone2022a}:
\begin{gather}\label{eq:1}
    \allowdisplaybreaks
    \begin{split}
    \allowdisplaybreaks
    \dfrac{\partial\Sigma}{\partial t}=\dfrac{3}{R}\dfrac{\partial}{\partial R}\left[R^{1/2}\dfrac{\partial}{\partial R}\left(\dfrac{\alpha_{\rm SS}\Sigma_{\rm g}c_{\rm s}^2}{\Omega_\mathrm{K}}R^{1/2}\right)\right]\\
    +\dfrac{3}{2R}\dfrac{\partial}{\partial R}\left(\dfrac{\alpha_{\rm DW}\Sigma_{\rm g}c_{\rm s}^2}{\Omega_{\rm K}}\right)\\
    -\dfrac{3\alpha_{\rm DW}\Sigma_{\rm g}c_{\rm s}^2}{4(\lambda-1)R^2\Omega_{\rm K}},
    \end{split}
\end{gather}
where we call $R$ the cylindrical radius and $t$ the time variable. 

The first term in Eq.~\ref{eq:1} describes how the total disc surface density $\Sigma$ varies as a consequence of viscous-diffusion processes. Here $\Sigma_{\rm g}$ is the gas surface density, and viscosity, $\nu=\alpha_{\rm SS}c_{\rm s}^2/\Omega_{\rm K}$, is written according to the \citet{Shakura&Sunyaev1973} prescription. $\alpha_{\rm SS}$ labels the angular momentum transport efficiency due to viscosity, $\Omega_{\rm K}$ is the Keplerian angular velocity, computed for a Solar mass star, and $c_{\rm s}$ is the locally-isothermal sound speed, which is determined as $c_{\rm s}\propto T^{1/2}$, where $T$ is the disc temperature. In our models, this is set by the stellar irradiation only: as we are interested in the evolution of disc sizes we neglect viscous heating, which is important only in the innermost disc regions. Hence $T$ is constant in time and decays as $R^{-1/2}$, with $T_0=88.23\,\mathrm{K}$ at $10\,\mathrm{au}$, tailored to a Solar mass star \citep{Kenyon&Hartmann1987,Chiang&Goldreich1997}. 

The second and third term in Eq.~\ref{eq:1} identify the contribution of magnetic winds to disc evolution. The second term describes the advection of disc gas due to angular momentum removal by magnetic torques associated with acceleration of the wind. In the ``expanded $\alpha$ framework'' of \citet{Tabone2022a}, this is parametrised similarly to the viscous term: as the viscous accretion rate, $\dot{M}_{\rm acc}^{\rm visc}$, is proportional to $\alpha_{\rm SS}$, so the wind-driven accretion rate, $\dot{M}_{\rm acc}^{\rm wind}$, is proportional to $\alpha_{\rm DW}$, making these two coefficients easy to compare. To first order, $\alpha_{\rm DW}$ is proportional to the disc magnetisation, generally denoted by the inverse of the thermal to magnetic pressure ratio in the disc mid-plane, $\beta_0$ \citep{Tabone2022a}. The third term identifies the mass loss rate in the wind, whose impact on disc evolution is described by the magnetic lever arm parameter, $\lambda$, defined as the ratio of the angular momentum carried away by the wind along a streamline and at its base \citep{Blandford&Payne1982}.

Eq.~\ref{eq:1} is solved using finite-differences methods on a grid made up of 4096 cells equally spaced in $R^{1/2}$, between $R_{\rm in}=10^{-2}\,{\rm au}$ and $R_{\rm out}=10^4\,{\rm au}$. In Appendix~\ref{app:1}, we show that our code exactly reproduces the analytical self-similar solutions of \citet{Tabone2022a}. As initial condition we consider a tapered power-law profile \citep{Lynden-Bell&Pringle1974}:
\begin{equation}\label{eq:2}
    \Sigma(R,t=0)=\dfrac{M_0}{2\pi R_0R}\exp\left(-\dfrac{R}{R_0}\right),
\end{equation}
where $M_0=0.1\,M_{\rm Sun}$ is the initial disc mass and $R_0$ is a characteristic radius, initially enclosing 63 per cent of $M_0$. A proper choice of the initial disc radius is particularly important in magnetic wind models, where gas sizes remain fixed with time \citep[e.g.,][]{Armitage2013,Tabone2022a}. Recent results from CALYPSO, \citep{Maury2019} and VANDAM, \citep{Tobin2020} suggest that disc sizes (estimated from dust emission) in Class 0/I sources are small, with only $\approx30\%$ of the targets being larger than $50\,{\rm au}$\footnote{It is unclear how dust and gas sizes are related in such young sources. However, Class~0 disc observations show that evidence of Keplerian rotation on scales larger than $50\,{\rm au}$ is rare (2/16 targets in the CO and SO line PDBI/CALYPSO observations of \citealt{Maret2020}).}. Yet, as Class 0/I discs are actively accreting material from their envelope, their observed sizes do not necessarily give strong constraints on the initial disc radius. For this reasons, following \citet{Rosotti2019_radii}, we set $R_0=10,\,30$ and $80\,{\rm au}$.

\paragraph*{Parameter space exploration} Our aim is to make predictions of how the observed disc sizes change as a function of the wind intensity. This is described by the parameter $\psi=\alpha_{\rm DW}/\alpha_{\rm SS}\approx\dot{M}_{\rm acc}^{\rm wind}/\dot{M}_{\rm acc}^{\rm visc}$, that quantifies the relative strength of the wind to viscous torque. While we can think of $\alpha_{\rm DW}$ as being dependent on $\beta_0$, the value of $\psi$ will depend on the configuration of the magnetic field (bipolar for MHD disc winds) and the disc micro-physics (that tells us if MRI can be triggered), but not necessarily on the disc magnetisation. When $\psi=0$, Eq.~\ref{eq:1} reduces to the viscous evolution equation, while when $\psi\rightarrow\infty$ magnetic winds dominates accretion. A fair comparison between models with different $\psi$ requires consideration of discs evolving on the same initial time scale:
\begin{equation}\label{eq:3}
    t_\mathrm{acc,0}=\dfrac{R_0^2}{3\alpha c_{\rm s}^2/\Omega_{\rm K}},
\end{equation}
where $\alpha=\alpha_{\rm SS}+\alpha_{\rm DW}$ is the total angular momentum transport efficiency in the disc. When $\psi=0$, $t_{\rm acc,0}$ is the initial viscous time scale, $t_{\nu,0}$.

In the following, we consider models with radially homogeneous $\alpha_{\rm SS}$ (resulting in $\nu\propto R$) and $\alpha_{\rm DW}$, for different values of $\psi$, corresponding to purely viscous evolution ($\psi=0$), the MHD wind dominated case ($\psi=10^4$), and a hybrid scenario ($\psi=1$). In a similar exercise in the viscous-only case, \citet{Rosotti2019_radii} considered $\alpha=10^{-4},\,10^{-3},\,10^{-2}$, encompassing the typical range bounded by hydro-dynamical instabilities to MRI, and $\alpha=0.025$, for illustrative purposes (and here motivated by the recent detection of high levels of turbulence in DM~Tau, \citealt{Flaherty2020}). The lower values of viscosity are also compatible with the results of \citet{Trapman2020}, who modelled CO disc sizes in Lupus, and those of \citet{Rosotti2019_fr} and \citet{Zormpas2022}, for the size-luminosity correlation. We retain the same values of $\alpha$ for an easier comparison of our results with those in the literature. This choice is also motivated by the currently available upper limits on the magnetic field strength \citep{Vlemmings2019,Harrison2021}, which cannot be used to significantly constrain $\alpha_{\rm DW}$ according to the present theoretical models. For reference, our choice of temperature implies that $t_{\rm acc,0}\approx 0.48\,{\rm Myr}$ at $R_0=10\,{\rm au}$ for $\alpha=10^{-3}$.

For each $\psi$ and $\alpha$ we also explore different values of the magnetic lever arm parameter, $\lambda=1.5,\,3$. These are justified by recent observations of disc molecular outflows (e.g., \citealt{Louvet2018,deValon2020,Booth2021}, where $1.5\lesssim\lambda\lesssim2.3$). We remark that this parameter does not set the disc evolution time scale (Eq.~\ref{eq:3}); instead it controls the amount of mass removed by the wind rather than accreted.

\citet{Tabone2022a} also considered models with a time-increasing $\alpha_{\rm DW}$, inversely proportional to the disc mass (corresponding to a locally constant $B_z$, rather than $\beta_0$). In a companion paper, \citet{Tabone2022b} also showed that this solution reproduces the correlation between proto-planetary disc accretion rates and masses inferred from (sub-)millimetre fluxes in Lupus \citep{Manara2016,Rosotti2017,Lodato2017}, as well as the decline of the disc fraction with cluster age \citep[e.g.,][]{Fedele2010}. For this reason, we take into account the non-constant $\alpha_{\rm DW}$ scenario as well. It is discussed in Section~\ref{sec:6}, because its results are very similar to the constant $\alpha_{\rm DW}$ case when $\psi\rightarrow\infty$, as long as the disc lives long enough.

\paragraph*{Dust evolution} The code implements the two population model of dust growth developed by \citet{Birnstiel2012} because it has the advantage of closely reproducing the results of full coagulation simulations \citep{Brauer2008,Birnstiel2010} at a much lower computational cost. In summary, two populations of solids are considered: one is made of small, monomer-sized grains ($a_0=10^{-5}\,{\rm cm}$, constant in time and space), the other is composed by large grains dominating the mass. Their size, the maximum grain size, $a_{\rm max}$, changes with time as the disc evolves and is set at any radius by the combination of growth (with maximally efficient sticking, $f_{\rm grow}=1$, \citealt{Booth&Owen2020}), fragmentation (assuming $u_{\rm f}=10\,{\rm m}\,{\rm s}^{-1}$ as velocity threshold for shattering, typical of ice-coated grains, e.g., \citealt{Gundlach&Blum2015}) and drift. 
In particular, after an initial phase of growth, $a_{\rm max}$ is determined by either the \textit{turbulent} fragmentation limit, $a_{\rm frag}\propto\Sigma_{\rm g}/\alpha_{\rm SS}$, or the drift limit, $a_{\rm drift}\propto\Sigma_{\rm d}$, whichever is the smallest \citep{Birnstiel2012}:
\begin{equation}\label{eq:4}
    a_{\rm max}=\min\left\{a_{\rm drift},\,a_{\rm frag},\,a_0\exp\left(t\epsilon\Omega_{\rm K}\right)\right\},
\end{equation}
where $\epsilon$ is the dust-to-gas ratio. From the previous expressions it is clear that the most important parameter in determining the grain size is the \textit{viscous} $\alpha_{\rm SS}$ parameter, because this sets the level of turbulent relative velocities that limit grain growth by fragmentation. Interestingly, the wind-dominated disc evolution scenario provides a physically-motivated application of previous works (e.g., \citealt{Pinilla2021}) studying dust evolution when the global angular momentum transport is decoupled from turbulence on a small scale. We refer to \citet{Birnstiel2012} for further details and to \citet{Booth2017} for the implementation.

Once the grain sizes are known, they are used to compute the dust velocities of both particle species as in \citet{Tanaka2005} and advect dust along the gas flow. To do so, the dust fraction is updated according to the one-fluid approach of \citet{Laibe&Price2014}, assuming that dust is initially distributed as the gas and 100 times less abundant \citep[as in the ISM, e.g.,][]{Bohlin1978}. The code allows to take into account the back reaction of the dust on the gas \citep[e.g.,][]{Dipierro2018,Garate2020}. However, this is likely unimportant given that the dust fraction decreases fast due to the efficient radial drift of solids \citep{Rosotti2019_radii}. 

In this paper dust entrainment in the wind is neglected based on previous works showing that only small grains can be efficiently removed in MHD winds \citep[e.g.,][]{Giacalone2019,Rodenkirch&Dullemond2022}. We discuss this assumption more extensively in Section~\ref{sec:6}.

Dust diffusion is also considered. The original code implements a diffusion coefficient $D=\nu/{\rm Sc}$, where ${\rm Sc}$ is the dust Schmidt number, computed as in \citet{Youdin&Lithwick2007}. To avoid non-smooth (numerical) features in the purely magnetic case, where viscosity is negligible, we prescribe $D=(\psi+1)\nu/{\rm Sc}$, so that there is a fixed relationship between angular momentum transport efficiency and diffusivity, regardless of the mechanism for angular momentum transport. Although we do not expect angular momentum transport due to magnetic winds to contribute to turbulent diffusion of solids, their mixing efficiency remains uncertain. Furthermore, the latter choice does not change our results on dust disc sizes.\\

\noindent As a summary of the previous paragraphs, the parameters used in our models are reported in Table~\ref{tab:1}. The values employed in the test case of Section~\ref{sec:3} are boxed.

\begin{table}
    \centering
    \begin{tabular}{l|l}
    \hline
        Parameter & Value \\
        \hline
        Torque ratio, $\psi$    & $0,\,1,\,10^4$\\[2.5pt]
        Lever arm parameter, $\lambda$ & $\boxed{3},\,1.5$\\[2.5pt]
        Accretion efficiency, $\alpha$  & $10^{-4},\,\boxed{10^{-3}},\,10^{-2},\,0.025$\\[2.5pt]
        Accretion time scale, $t_{\rm acc}$ & $4.872,\,\boxed{0.487},\,0.048,\,0.019\,{\rm Myr}$\\[2.5pt]
        Initial disc radius, $R_0$     & $\boxed{10},\,30,\,80\,{\rm au}$\\[2.5pt]
        Disc mass, $M_0$     & $0.1\,M_{\rm Sun}$\\[2.5pt]
        Stellar mass, $M_*$     & $1\,M_{\rm Sun}$\\[2.5pt]
        Reference temperature, $T_0$     & $88.23\,{\rm K}$\\[2.5pt]
        Dust material density, $\rho_{\rm s}$ & $1\,{\rm g}\,{\rm cm}^{-3}$\\[2.5pt]
        Fragmentation velocity, $u_{\rm f}$    & $10\,{\rm m}\,{\rm s}^{-1}$\\[2.5pt]
        Initial grain size, $a_0$     & $10^{-5}\,{\rm cm}$\\[2.5pt]
        Initial dust-to-gas ratio, $\epsilon$& $10^{-2}$\\[2.5pt]
        Coagulation efficiency, $f_{\rm grow}$ & 1 \\
    \hline
    \end{tabular}
    \caption{Summary of the model parameters, references can be found in the text. When a range of parameters is explored, the values used in our test case in Section~\ref{sec:3} are boxed.}
    \label{tab:1}
\end{table}

\paragraph*{Post-processing and disc size determination} The outputs of our simulations are post-processed to generate synthetic observations at ALMA wavelengths. To do so we first compute the surface brightness radial profile, $S_\mathrm{b}$, at $\nu\approx352.7\,{\rm GHz}$ ($\approx 0.85\,{\rm mm}$, ALMA Band~7) and $\nu\approx96.7\,{\rm GHz}$ ($\approx 3.10\,{\rm mm}$, ALMA Band~3) as:
\begin{equation}\label{eq:5}
    S_{\rm b}(R)=B_\nu(T)\bigl\{1-\exp{(-\kappa_\nu\Sigma_{\rm d})}\bigr\},
\end{equation}
where $\Sigma_\text{d}$ is the dust surface density, $B_\nu$ is the black body radiation spectrum at temperature $T$ and $\kappa_\nu$ is the dust (absorption) opacity, computed as in \citet{Rosotti2019_radii}, following \citet{Tazzari2016}. We refer to \citet{Rosotti2019_fr} for comments on this choice of opacity. For simplicity, we assume discs to be face-on, which is a fair approximation within a factor $\langle\cos i\rangle=\pi/4\approx0.8$ from the average disc inclination on the sky. 

In this paper we consider smooth discs (i.e., continuous, without sub-structures), where defining a characteristic scale radius is arbitrary. A possible solution naturally comes from our parametrisation of the initial conditions in Eq.~\ref{eq:2}, where the scale radius, $R_0$, allows to infer how big a disc is. Although this definition naively applies only at the very beginning of the disc life, it can be extended to later times and regardless of the shape of the disc surface density. For example, following \citep{Rosotti2019_radii}, this can be done cumulatively defining as disc radius the location in the disc enclosing a given fraction of the total disc mass. For consistency with the physical interpretation of $R_0$, we define the \textit{disc mass radius}, $R_{\rm 63,mass}$, as the disc size enclosing 63 per cent of the total gas or dust mass, whichever tracer is considered. As this metric is not observationally accessible, we also define an observational \textit{disc flux size} as the radius enclosing a given fraction of the disc luminosity, $R_{x,{\rm flux}}$. Typical values of $x$ employed in the literature are 68 and 95 per cent \citep[e.g.,][]{Tripathi2017,Andrews2018,Manara2019,Hendler2020,Sanchis2021,Tazzari2021_models}. \citet{Rosotti2019_radii} suggested that $R_{\rm 68,flux}$ indicates the location in the disc where grains are about the size of the observational wavelength (or the \textit{opacity cliff}, as named by \citealt{Rosotti2019_radii}). $R_{\rm 95,flux}$, instead, is likely a proxy for the outer disc radius, observationally set by the survey sensitivity.

\subsection{Observational sample} 
To test the viscous and magnetic evolution model predictions, we compare our synthetic disc sizes with those observationally-inferred from Lupus \citep{Ansdell2016}, Chamaeleon~I \citep{Pascucci2016,Long2018_cham} and Upper~Sco \citep{Barenfeld2016} ALMA data in Band~7. We mainly refer to the work of \citet{Hendler2020}, who computed dust disc sizes homogeneously among these regions, and use the the PPVII Chapter table of \citealt{Manara2022}, who adopted Gaia-EDR3 distances \citep{GAIAeDR3}. We choose these star-forming regions because they are among the best studied nearby ones, with high levels of completeness and disc detection fractions. Moreover, they span a wide age range: Lupus ($\approx1\,{\rm to}\,3\,{\rm Myr}$, \citealt{Comeron2008}), Chameleon~I ($\approx2\,{\rm to}\,3\,{\rm Myr}$, \citealt{Luhman2008}), Upper Sco ($\approx5\,{\rm to}\,10\,{\rm Myr}$, \citealt{Preibisch2002}). This is ideal to test our model predictions over a long time interval. In Table~\ref{tab:2} we summarise the main properties of the data-sets taken into account.

\begin{table}
    \centering
    \begin{tabular}{c|c|c|c|c}
    \hline
    Region       & Observed & Measured & Resolved & $R_{\rm68,flux}$ \\
    \hline
    Lupus        & 89  & 50 & 30 & $46.82^{+45.52}_{-30.88}$\\[2.5pt]
    Chamaeleon~I & 93  & 58 & 33 & $32.49^{+35.70}_{-8.81}$\\[2.5pt]
    Upper Sco    & 106 & 44 & 22 & $25.05^{+15.58}_{-12.37}$ \\
    \hline
    \end{tabular}
    \caption{Size properties of the Lupus \citep{Ansdell2016}, Chamaeleon~I \citep{Pascucci2016,Long2018_cham} and Upper Sco \citep{Barenfeld2016} samples taken into account: number of observed discs, discs where sizes were measured and resolved, median and $1\sigma$ uncertainty of $R_{\rm68,flux}$ for resolved discs only. Data on measured and resolved discs are from \citet{Hendler2020}. Note that the Gaia-corrected median distances for resolved discs are in line with those reported by \citet{Hendler2020}.}
    \label{tab:2}
\end{table}

We also compare the behaviour of our models with Lupus and Upper Sco data in the disc size \textit{vs} luminosity plane. We use the sizes and fluxes computed by \citet{Hendler2020} from ALMA Band~7 data and those derived by \citet{Tazzari2021_models} from ALMA Band~3 \citep{Tazzari2021_obsv} data in sub-sample of Lupus discs.

\section{Results in a test case}\label{sec:3}
We discuss the impact of magnetic winds with different mass-loss rates on dust evolution in a test case with $\alpha=10^{-3}$ and $R_0=10\,{\rm au}$. Then the evolution of the mass radius and flux radius is detailed.

\begin{figure*}
    \centering
    \includegraphics[width=\textwidth]{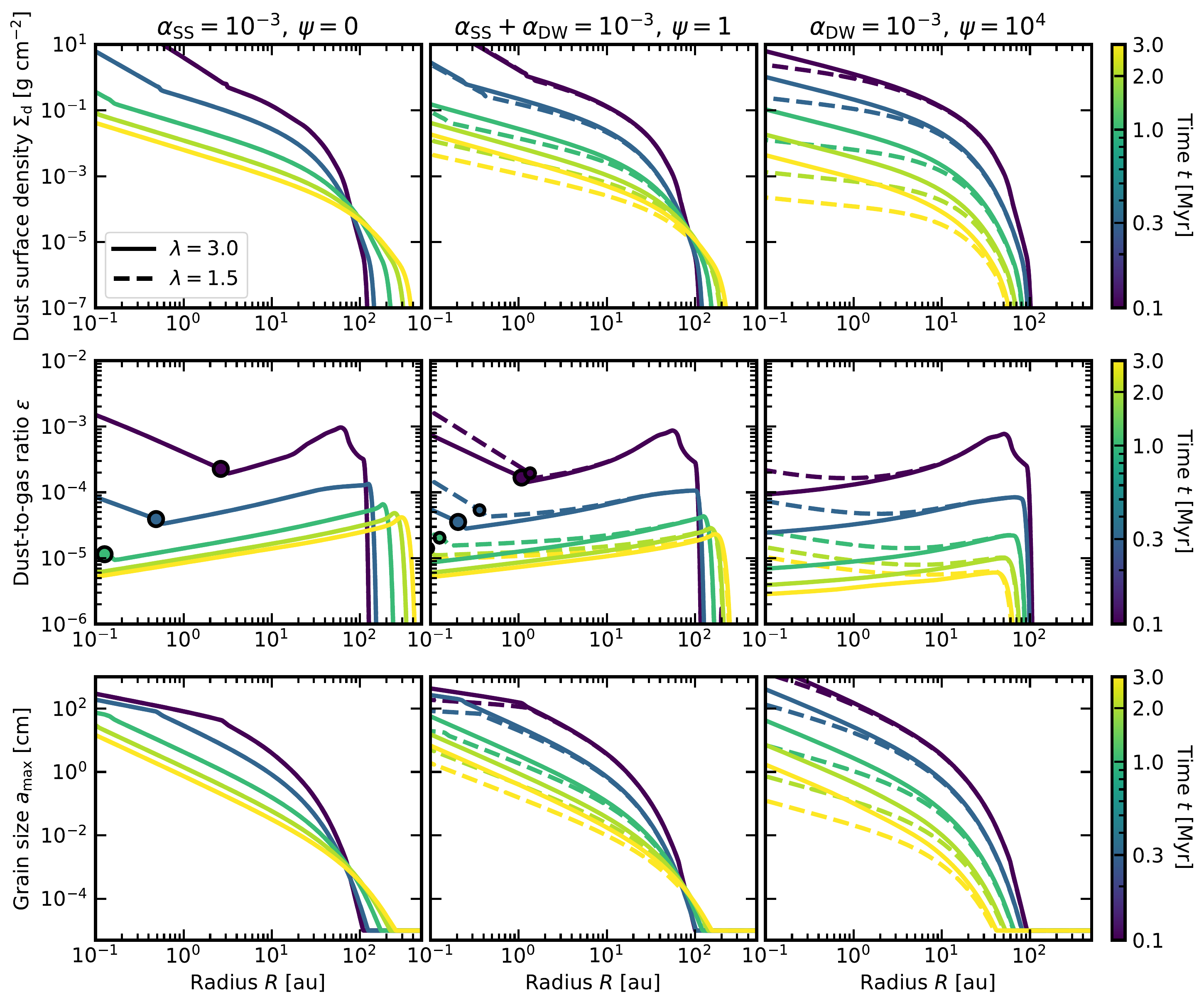}
    \caption{\textbf{Upper panels:} Dust surface density radial distribution for different disc evolution models in the test case ($\alpha=10^{-3},\,R_0=10\,{\rm au}$). From left to right: viscous evolution ($\psi=0$), hybrid case ($\psi=1$), magnetic wind evolution ($\psi=10^4$). Solid lines are used for moderate mass loss rates ($\lambda=3$), while dashed lines for less-efficient winds ($\lambda=1.5$). \textbf{Middle panels:} Dust-to-gas ratio radial distribution. Dots mark the transition between fragmentation-dominated and drift-dominated dust regime, larger for $\lambda=3$ and smaller for $\lambda=1.5$. \textbf{Bottom panels:} Maximum grain size radial distribution.}
    \label{fig:3.1}
\end{figure*}

\subsection{Dust evolution in the test case}
Fig.~\ref{fig:3.1} compares dust evolution profiles at different times in viscous and magnetic wind models. The radial profiles of the dust surface density, $\Sigma_{\rm d}$, dust-to-gas ratio, $\epsilon$, and maximum grain size, $a_{\rm max}$, are displayed from top to bottom and each column highlights models with different torque ratios: from left to right, the viscous evolution case ($\psi=0$, bench-marked against \citealt{Rosotti2019_radii} results), the hybrid scenario ($\psi=1$) and the purely magnetic wind case ($\psi=10^4$).

Let us first focus on the viscous case (left panel): the main features of dust evolution were already discussed in previous works \citep[e.g.,][]{Birnstiel2012,Rosotti2019_radii} and are briefly summarised hereafter:
\begin{itemize}
    \item dust depletes very fast and by $t=3\,{\rm Myr}$ the dust-to-gas ratio falls below $\epsilon=10^{-5}$;
    \item initially dust growth is limited by fragmentation in the inner disc and by dust drift in the outer disc: the transition between the two regimes is identified by the kink in the dust-to-gas ratio profiles (see dots in Fig.~\ref{fig:3.1}). The drift-dominated regime encompasses larger disc regions with time;
    \item the dust surface density displays a sharp outer edge at any time \citep{Birnstiel&Andrews2014}. This is determined by the abrupt change in dust velocity at these locations.
\end{itemize}

We can discuss now how these characteristics change when discs evolve under the effect of magnetic winds. As can be seen from Fig.~\ref{fig:3.1}, in the purely magnetic wind case (solid lines in the right panel), dust growth is limited only by radial drift and no kink in the dust-to-gas ratio can be seen. This happens because, at fixed $\alpha$, the magnetic wind models have lower levels of viscosity. Therefore no \textit{turbulent} fragmentation takes place, and the disc is drift-dominated throughout. What is more, discs also lose solids faster: dust masses are 50 times smaller than in the purely viscous case, after $t=3\,{\rm Myr}$, when $\psi=10^{-4}$. A possible explanation is as follows. In the absence of winds, because of viscous spreading, small grains well coupled with the gas can migrate outwards. Those grains make up a reservoir of solids that grow, decouple from gas and eventually drift on very long time scales, replenishing the inner disc with dust at later times. In the purely magnetic wind case, instead, no outward diffusion is possible and the solids sink (faster) on the star as dust velocities are directed inwards at any radius, causing the disc outer edge to move accordingly. Similar results were described by \citet{Sellek2020} in the case of discs evolving under the effect of external photo-evaporation and by \citet{Zagaria2021_theory} for discs in binary systems. 

In the hybrid case (solid lines in the central panel), an intermediate behaviour is present. Because of the reduced contribution of viscosity to the angular momentum transport, discs become drift-dominated at a factor of 2 to 3 smaller radii, but fragmentation is still important in the inner disc. Dust masses are 3 times smaller than in the purely viscous case, after $t=3\,{\rm Myr}$, but not as low as in the purely magnetic wind case because a small amount of spreading is present.

In Fig.~\ref{fig:3.1} we also consider the effect that changing the lever arm parameter has on dust evolution. The solid lines discussed so far refer to the $\lambda=3$ case, while the dashed lines are used for models with $\lambda=1.5$. In this case, when discs are dominated by MHD winds the dust surface density and the maximum grain size rise less strongly in the inner disc because the gas surface density is shallower in the same regions ($\Sigma_{\rm d}^2\propto\Sigma_{\rm g}/\Omega_{\rm K}R^2$ in the drift-dominated regime, and $a_{\rm drift}\propto\Sigma_{\rm d}$, see \citealt{Birnstiel2012}). In the hybrid model, the transition between fragmentation-dominated and drift-dominated regions of the disc moves to larger radii. This can be explained on account of the higher dust-to-gas ratio in the inner disc ($a_\mathrm{drift}/a_{\rm frag}\propto\epsilon$), motivated by the larger gas loss with respect to the $\lambda=3$ case (see also discussion in \citealt{Tabone2022b}). Finally, although the (locally) larger dust-to-gas ratio may hint at the expectation that more dust is retained when $\lambda=1.5$, these models lose two to three times more solids than in the case of $\lambda=3$. This is due to the faster velocity of grains, that decouple earlier from the gas because of the lower gas surface density\footnote{This dominates over the change of surface density slope in the inner disc, which has the opposite effect of reducing the dust velocity, that is proportional to the local pressure gradient \citep{Brauer2008}.}.

\subsection{Dust disc sizes in the test case}
\begin{figure*}
    \centering
    \includegraphics[width=\textwidth]{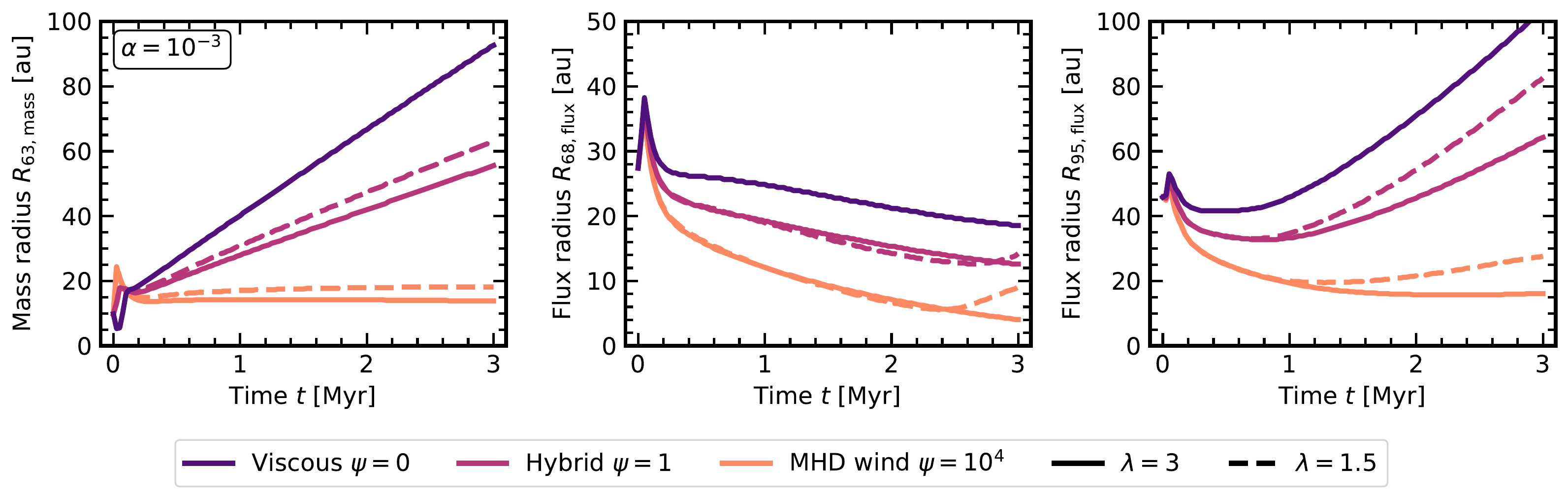}
    \caption{\textbf{From left to right:} dust mass radius and 68 and 95 per cent flux sizes as a function of the disc age for different evolution models. Solid and dashed lines display $\lambda=3$ and $1.5$ models, respectively.}
    \label{fig:3.2}
\end{figure*}

Our models are evolved from 0 to $3\,{\rm Myr}$. In the following, we consider how their characteristic sizes change after every $5\times10^4\,{\rm yr}$ snapshot. In Fig.~\ref{fig:3.2} the dust mass radius and the 68 and 95 per cent flux sizes at $352.7\,{\rm GHz}$ (i.e. $0.85\,{\rm mm}$, ALMA Band 7), are plotted as a function of the disc age for different values of $\psi$. Solid and dashed lines display $\lambda=3$ and $1.5$ models, respectively.

Let us consider the $\lambda=3$ case first. In the left panel of Fig.~\ref{fig:3.2}, we can see that magnetic winds substantially affect the mass radius evolution. While in the viscous model, $R_{\rm63,mass}$ increases with time, this is not always true when magnetic winds are considered. In particular, the higher $\psi$ is, the less the mass radius grows. Eventually, in the purely magnetic wind case, after an initial transient determined by the removal of large dust \citep{Rosotti2019_radii}, $R_{\rm63,mass}$ remains steady. This is due to the absence of viscous spreading that allows for small, well-coupled grains to move to larger and larger radii, increasing the mass radius. In fact, by $t=3\,{\rm Myr}$ the purely MHD wind model is smaller than the purely viscous one by a factor of five. 

While the behaviour of the mass radius is insensitive to its definition ($R_{x,{\rm mass}}$, with $x\in[50,100]$), this is not true for the flux radius. On the one hand, as is shown in the central panel of Fig.~\ref{fig:3.2}, $R_{\rm68,flux}$ always decreases with time, regardless of $\psi$. On the other hand, in the right panel of Fig.~\ref{fig:3.2} we can see that $R_{\rm95,flux}$ closely follows the behaviour of the mass radius: it increases the most in the viscous case and remains steady in the purely magnetic wind model.

Moving to the $\lambda=1.5$ case, it can be seen from Fig.~\ref{fig:3.2} that the absolute values of the disc sizes are sensitive to the magnetic lever arm parameter. Specifically, they are larger for smaller values of $\lambda$, as can be understood by the less steep dust surface density in this case. Nevertheless, variations are small, usually within 20 to 30 per cent. 
Most importantly, the increasing/decreasing trend of the disc sizes is not influenced by the lever arm parameter\footnote{In the purely magnetic wind models, at late times $R_{\rm68,flux}$ and $R_{\rm95,flux}$ increase when $\lambda=1.5$. This is because the brighter inner disc shrinks so much that the fainter regions out of the \textit{opacity cliff} become significant to the total disc luminosity \citep{Sellek2020,Toci2021}. This takes place earlier for $\lambda=1.5$ than for $\lambda=3$ because of the faster removal of the grains with size beyond the cliff.}. This is why in the following we will only consider the $\lambda=3$ case and move to Appendix~\ref{app:3} a discussion on how our results change when $\lambda=1.5$.

\section{Parameter space exploration}\label{sec:4}
In this Section we discuss the behaviour of the mass and flux radius for different values of $\psi$, as a function of the initial disc size, $R_0$, and angular momentum transport efficiency, $\alpha$.

\subsection{The evolution of the mass radius}
\begin{figure*}
    \centering
    \includegraphics[width=\textwidth]{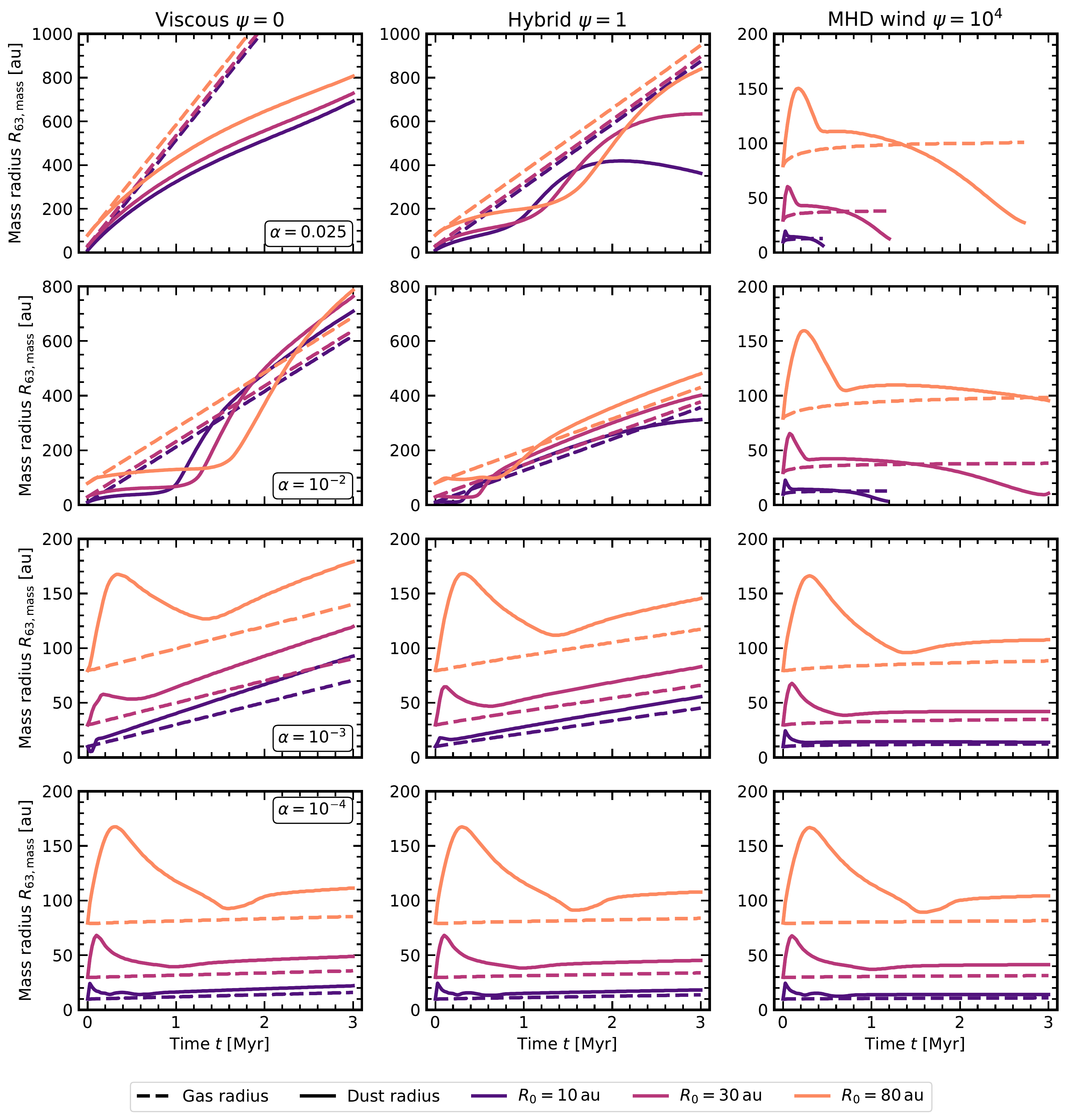}
    \caption{Mass radius for gas (dashed lines) and dust (solid lines) time evolution as a function of $\alpha$ (row by row) and $\psi$ (column by column), for $\lambda=3$.}
    \label{fig:4.1}
\end{figure*}

Fig.~\ref{fig:4.1} displays the evolution of the mass radius for different initial parameters. In each sub-plot solid lines are used for the dust, dashed lines for the gas, colour-coded according to the initial disc radius, $R_0$. Different values of $\alpha$ and $\psi$ are used, moving from the top to the bottom along a column and from the left to the right along a row, respectively.

\paragraph*{Dependence on the initial radius} Let us first focus on the viscous evolution models ($\psi=0$) with $\alpha=10^{-3}$. As already discussed in the previous Section and in \citet{Rosotti2019_radii}, $R_{\rm63,mass}$ increases with time, and is larger for a larger $R_0$. When the disc scale radius is big enough, an initial transient (peak) can be seen. This is an effect of grain growth and radial drift. (i) First, the large and fast growing particles in the inner disc are removed ($R_{\rm63,mass}$ increases). (ii) Then, the grains drift as they grow also in the outer disc regions ($R_{\rm63,mass}$ decreases). (iii) Finally, when only the smallest, well-coupled grains are left, the mass radius expands again. As discussed in \citet{Rosotti2019_radii}, this effect is more evident for a larger $R_0$ because of the longer evolution time scales, implying a slower growth, drift and outward diffusion of solids at large radii. 

When magnetic winds dominate disc evolution ($\psi>0$), similar behaviours with $R_0$ can be observed. However, $R_{\rm63,mass}$ increases less or remains steady because of the reduced viscosity contribution to angular momentum transport.

\paragraph*{Dependence on $\alpha$} Let us first focus on the viscous evolution models ($\psi=0$). When $\alpha=10^{-4}$, viscosity is so low that no late time expansion can be observed, regardless of $R_0$. For this reason, no differences can be noticed with the magnetic wind cases.

For a larger $\alpha$, the models undergo two expansion phases \citep{Rosotti2019_radii}. For $\alpha=10^{-2}$, the mass radius initially grows slowly, then substantially expands because grains become smaller and well coupled with the gas (corresponding to the bulk of the disc transitioning from the fragmentation-dominated to the drift-dominated regime). In this case, $R_{\rm63,mass}$ can be larger for a smaller initial radius, because grains become drift-dominated earlier. Instead, when $\alpha=0.025$, after expanding with the gas, the dust radius grows less than the gas radius because of the small grains in the outer disc decoupling from the gas and drifting inwards. 

In the purely magnetic wind models ($\psi=10^4$) the mass radius follows a completely different evolutionary path. After an initial transient, that can be motivated as in the previous paragraphs, $R_{\rm63,mass}$ stays constant or decreases, and eventually plummets, because of the fast drift of solids. This process takes place at different characteristic times: earlier for a larger $\alpha$ (because of the faster evolution) and a smaller $R_0$ (because of the faster dust removal). For $\alpha=10^{-2}$ and 0.025, we decided to plot the mass radius evolution up to the time step when all the large grains are removed as the disc would be dispersed by this stage (when the track is ended the highest measured dust mass in our models is $\approx4\times10^{-5}\,M_{\rm Jup}$, about the mass of the Moon, and generally $10^3$ times smaller). $R_{\rm63,mass}$ then rapidly increases and finally attains the same value as in the gas (not shown).

Finally, in the hybrid models ($\psi=1$), for both $\alpha=10^{-2}$ and 0.025, the behaviour of $R_{\rm63,mass}$ resembles the viscous case with $\alpha=10^{-2}$, as can be expected by the reduced turbulence in hybrid models. However, the late stages of evolution are characterised by a moderate disc shrinkage as the small particles in the outer disc decouple from the gas and drift inwards. For lower values of $\alpha$ an intermediate behaviour between the viscous and purely MHD wind case can be seen.

\paragraph*{Dust-to-gas size ratio} Finally, we comment on the relationship between dust and gas radii. In the viscous scenario gas sizes always expand, while in the wind-dominated case they remain steady \citep{Armitage2013,Tabone2022a}. Dust disc sizes are larger or smaller than the gas ones depending on the steepness of the dust surface density with respect to the gas surface density\footnote{Our result of dust disc sizes being larger than gas disc sizes can be counter-intuitive. Indeed, the latter are generally predicted \citep{Facchini2017,Trapman2019,Toci2021} and observed (\citealt{Ansdell2018,Facchini2019,Andrews2020,Sanchis2021,Miotello2022}) to be larger than the former. However, our results are not necessarily in tension with the literature, because they are based on the total gas \textit{mass}, rather than the CO \textit{flux}. A proper comparison with the observations would require modelling CO emission, which is beyond the aims of this paper.}. The latter is given by $-1+\xi$ \citep{Tabone2022a}, where $\xi=d\log\dot{M}_{\rm acc}^{\rm wind}/d\log R$ \citep{Ferreira&Pelletier1995} is known as the ejection index. The former is mainly determined by the mechanisms that set the grain size in the disc. Adopting the same method of \citet{Birnstiel2012}, but see also the analytical solutions of \citet{Birnstiel&Andrews2014}, it can be shown that in the fragmentation-dominated regime $\Sigma_{\rm d}\propto R^{-3/2}$, steeper than the gas, while in the drift-dominated one $\Sigma_{\rm d}\propto R^{(-3+2\xi)/4}$, which is always less steep than the gas for $\lambda=3$.\\

\noindent To sum up, in the viscous evolution models the dust mass radius always expands as long as viscosity is large enough ($\alpha\gtrsim10^{-3}$). Generally speaking, larger initial sizes and a larger $\alpha$ lead to larger discs. In the purely magnetic wind scenario, instead, the dust mass radius stays constant when the rate of angular momentum extraction in the wind is low ($\alpha\lesssim10^{-3}$). However, for stronger winds, $R_{\rm63,mass}$ decreases and solids rapidly disperse. 

\subsection{The evolution of the flux radius}
\begin{figure*}
    \centering
    \includegraphics[width=\textwidth]{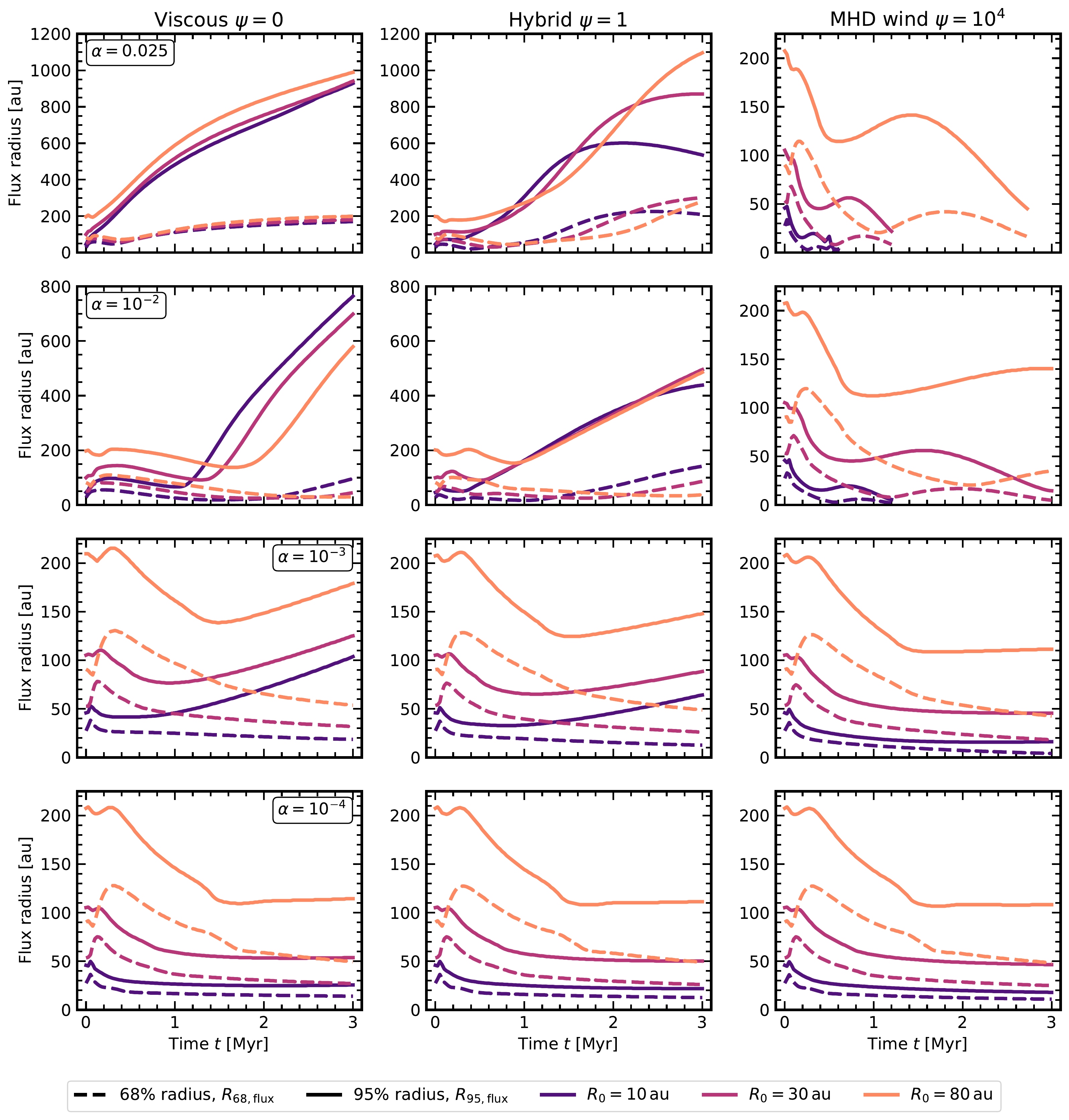}
    \caption{68 per cent (dashed lines) and 95 per cent (solid lines) flux radius time evolution as a function of $\alpha$ (row by row) and $\psi$ (column by column), for $\lambda=3$.}
    \label{fig:4.2}
\end{figure*}

In Fig.~\ref{fig:4.2} the evolution of the 68 and 95 per cent flux radius at $352.7\,{\rm GHz}$ (i.e. $0.85\,{\rm mm}$, ALMA Band 7), is shown in dashed and solid lines, respectively.

Regardless of $\psi$, for $\alpha\lesssim10^{-3}$ the 68 per cent flux radius decreases with time because the grains with the largest opacity ($a_{\rm max}\approx0.02\,{\rm cm}$, corresponding to the \textit{opacity cliff} size) in the inner disc drift towards the star. Instead, when $\alpha\gtrsim10^{-2}$, after an initial shrinking phase, $R_{\rm 68,flux}$ can get larger with time in the viscous and hybrid cases. This happens because the 68 per cent flux radius is not tracing the disc region where $a_{\rm max}$ is about the observational wavelength. In fact, the contribution of small grains ($a_{\rm max}\lesssim0.02\,{\rm cm}$, below the \textit{opacity cliff}) to the disc brightness becomes important in this case, because the larger particles were already removed. Moreover, $R_{\rm 68,flux}$ has a similar time-dependence as the mass radius\footnote{Once a disc is mainly made of small grains, it is reasonable to assume that it is optically thin. Then $S_{\rm b}\propto T\kappa_\nu\Sigma_{\rm d}$, and the surface brightness changes with time as the dust surface density does, because very small grains have a uniform opacity and the temperature does not vary with time in our models.}. 

The 95 per cent flux radius closely follows the behaviour of the mass radius. When angular momentum transport in negligible ($\alpha=10^{-4}$), after an initial shrinking phase due to the brightest grains drifting inwards (corresponding to the transient in the mass radius and enhanced by opacity effects), it remains steady. A similar initial decrease can be observed for $\alpha=10^{-3}$, with $R_{\rm 95,flux}$ eventually growing in the viscous and hybrid models and remaining steady in the purely magnetic wind case. When $\alpha\gtrsim0.01$, no shrinking is observed in the viscous and hybrid cases, because dust and gas are well coupled. As a consequence, the 95 per cent flux radius increases, until the small grains in the outer disc start drifting efficiently (e.g., $\psi=1,\,\alpha=0.025,\,R_0=10\,{\rm au}$, where $\Sigma_{\rm d}$ drops and the solids become drift-dominated). On the contrary, in the purely magnetic wind scenario, the 95 per cent flux radius never grows as expected from the absence of outwards disc spreading. 

In our experiments we also considered different definitions of the observed dust disc sizes, $R_{x,{\rm flux}}$, using intermediate flux percentages, $x$, between 68 and 95. However, these quantities are not useful to distinguish MHD wind and viscous evolution, because their ability to recover viscous expansion depends on $\alpha$.

To sum up, radial drift leads the central brightest region of the disc to shrink, regardless of the evolutionary scenario. Nevertheless, when sufficiently large fractions of the dust flux are used to define disc sizes, we recover the same behaviour of the mass radius. This is due to the faint outer disc expanding in viscous and hybrid models. Thus the 95 per cent flux radius is a promising proxy to distinguish viscous/hybrid evolution from the purely MHD wind one.

\section{Observational consequences}\label{sec:5}
In this Section we discuss our model prediction for the evolution of observed dust disc sizes and the size-luminosity correlation. Then, the results of our synthetic observations are compared with real data in Lupus, Chamaeleon~I and Upper Sco.

\subsection{Can observations discriminate between viscous and wind-driven evolution?}
Although the 95 per cent flux radius is a promising proxy for viscous expansion as opposed to magnetic wind shrinking or stalling, it was already highlighted by \citet{Rosotti2019_radii} that measurements of viscous spreading are expected to be challenging even for ALMA.

\begin{figure*}
    \centering
    \includegraphics[width=\textwidth]{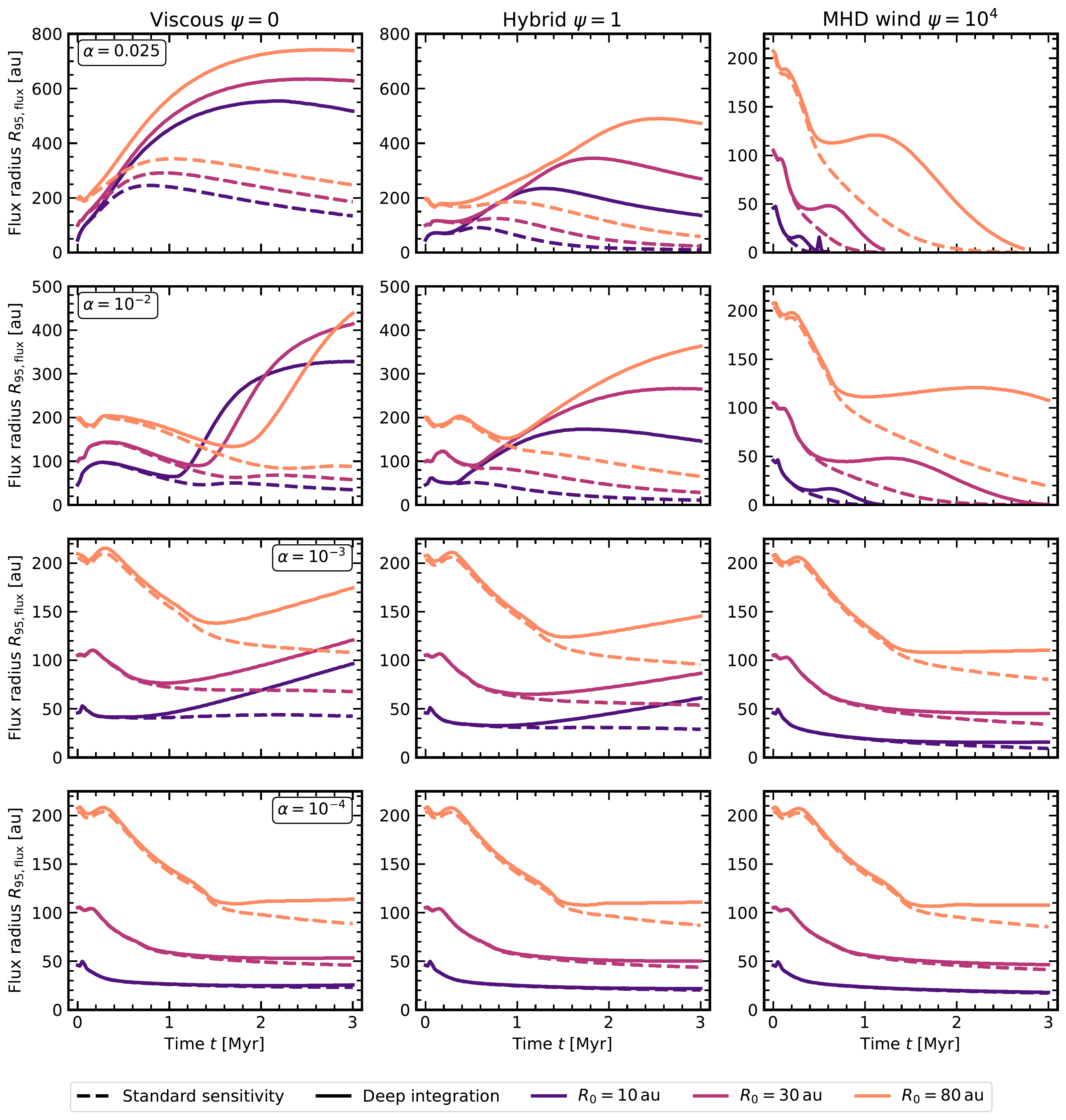}
    \caption{95 per cent flux radius time evolution as a function of $\alpha$ (row by row) and $\psi$ (column by column), for $\lambda=3$. The current survey case and the deep sensitivity scenario are plotted as dashed and solid lines, respectively.}
    \label{fig:5.1}
\end{figure*}

Current surveys are not deep enough to distinguish between viscous and wind-driven evolution. To motivate this statement, following the argument of \citet{Rosotti2019_radii}, we repeat our calculation of the flux radius neglecting dust emission falling below the sensitivity threshold given by $S_{\rm b,0.85mm}\approx6\times10^7\,{\rm Jy}\,{\rm sr}^{-1}$, corresponding to $\approx2\,{\rm min}$ integration time and $0.3\,{\rm arcsec}$ angular resolution observations, similar to those of \citet{Ansdell2016,Barenfeld2016,Pascucci2016}. The new sizes are shown in Fig.~\ref{fig:5.1} as dashed lines. $R_{\rm95,flux}$ always decreases regardless of the model parameters, making it impossible to distinguish between viscous and magnetic wind models. Noticeably, however, purely magnetic wind models with $\alpha\gtrsim10^{-2}$ would be already dispersed by this time, reducing the parameter space for comparing dust disc sizes among different disc evolution models to $\alpha\approx10^{-3}\text{ to }10^{-4}$.

As shown by \citet{Rosotti2019_radii}, deeper observations (e.g., with a sensitivity threshold of $S_{\rm b,0.85mm}\approx1\times10^6\,{\rm Jy}\,{\rm sr}^{-1}$) can approximately recover the theoretical values with no sensitivity cut, hence the different evolutionary trend for dust disc sizes in the viscous and magnetic case. This is displayed in Fig.~\ref{fig:5.1} in solid lines. However, achieving these high sensitivity is possible either drastically reducing angular resolution ($1\,{\rm arcsec}$ for $1\,{\rm hr}$ integration time) or through prohibitively long surveys ($0.67\,{\rm arcsec}$ for $5\,{\rm hr}$ integration time), as discussed in \citet{Rosotti2019_radii}. In the first case, which is the most feasible for observations involving more than one source, angular resolution would be enough only for viscous and hybrid models with $\alpha\geq10^{-2}$ to be resolved at the distance of nearby star-forming regions ($d\approx140\,{\rm au}$). In fact, these are the only models reaching dust disc sizes of more than $200\,{\rm au}$ after $1\,{\rm Myr}$ of evolution. Magnetic wind models, instead, would either disperse or be too small to be resolved. This suggests that very compact/unresolved discs (making up between 50 and 60 per cent of the Lupus population, according to \citealt{Miotello2021}), detected with very high sensitivity, could be evolving under the effect of magnetic winds\footnote{In the case of Lupus or Chamaeleon external processes reducing disc sizes such as photo-evaporation and tidal truncation due to fly-bys can be neglected.}. However, at the limited sensitivity of current surveys, it is also possible that those discs looking compact are instead larger and faint in their outer regions.

\begin{figure*}
    \centering
    \includegraphics[width=\textwidth]{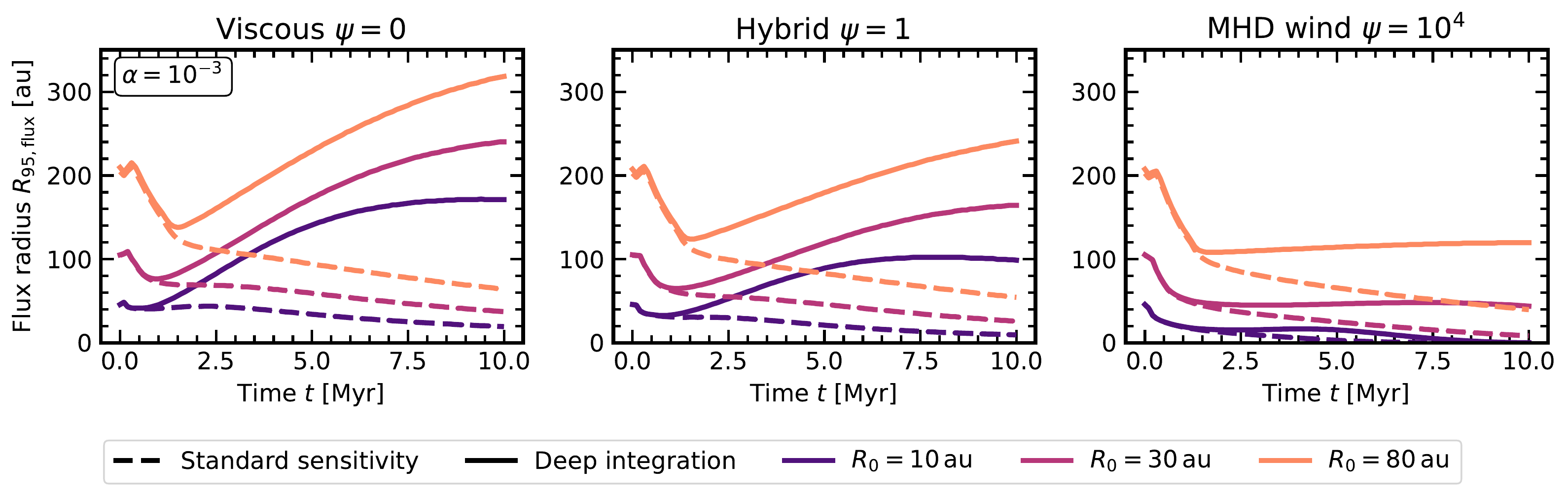}
    \caption{Same as in Fig.~\ref{fig:5.1} for models with $\alpha=10^{-3}$ evolving up to $10\,{\rm Myr}$: deep integration is required to discriminate wind-driven and turbulent accretion.}
    \label{fig:5.2}
\end{figure*}

The case of $\alpha=10^{-3}$ shows the most striking differences between viscous and (non-dispersing) magnetic wind models. However, these discs would perhaps be too small to be resolved by ALMA in the configuration with high sensitivity and shortest integration time discussed in the previous paragraph. On top of this, the initial decrease in $R_{\rm 95,flux}$ would make it difficult to assess if viscous expansion is actually taking place \citep{Rosotti2019_radii}. A possible solution to this problem is provided by considering the evolution of disc sizes between star-forming regions with a larger age difference.

We evolve our models with $\alpha=10^{-3}$ from 0 to $10\,{\rm Myr}$ and consider how their characteristic sizes change after every $1\times10^5\,{\rm yr}$ snapshot. In Fig.~\ref{fig:5.2} the same exercise as in Fig.~\ref{fig:5.1} is repeated for these longer-lived models. 
As can be seen, viscous expansion can be recovered if observations are deep enough. On the contrary, wind-dominated models have constant sizes. Furthermore, these differences are still observable considering disc sizes enclosing down to 85 per cent of the disc luminosity, which would be more observationally feasible and lead to more robust results. 
However, models of long-lived discs require a proper treatment of late-time dispersal (e.g., photo-evaporation), that is expected to complicate this picture. We refer to Section~\ref{sec:6} for a proper discussion.

In summary, current observations cannot distinguish between viscously evolving or wind-dominated disc models. In the case of deeper observations, viscous/hybrid models are characterised by dust disc sizes increasing over time, while purely MHD wind ones either shrink and disperse or stay the same. Even in this case, data sets with large age differences are required for a fruitful comparison.

\subsection{Comparison between the observationally-inferred dust disc sizes and our model predictions}
We compare dust disc sizes inferred from synthetic observations with those homogeneously computed by \citet{Hendler2020} and \citet{Manara2022}. We restrict to the case of $\alpha=10^{-3}\text{ and }10^{-4}$, because only in these cases purely magnetic wind models retain enough dust by $10\,{\rm Myr}$. Note that, this choice is also in line with the observed mass accretion rates in T~Tauri discs \citep[e.g.,][]{Lodato2017,Sellek2020_acc}. Then, the median $R_{\rm68,flux}$ is computed over the time corresponding to the average age of each star-forming region, for different values of the initial disc radius and $\psi$, neglecting dust emission below $S_{\rm b,0.85mm}\approx6\times10^7\,{\rm Jy}\,{\rm sr}^{-1}$.

\begin{figure*}
    \centering
    \includegraphics[width=\textwidth]{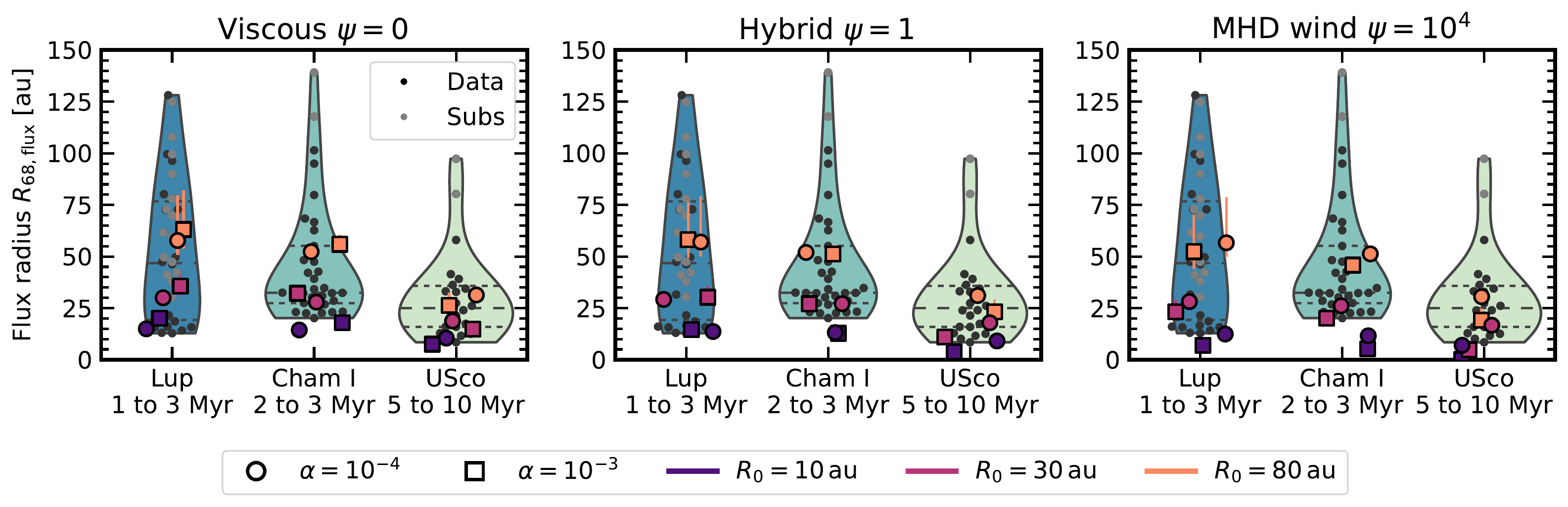}
    \caption{68 per cent flux radius for Lupus (Lup), Chamaeleon~I (Cham~I) and Upper Sco (USco) discs in black (and grey when sub-structured). Dashed and dotted lines in the violin plot indicate the median of the data and their 16th and 84th percentiles. Large squares and dots display the median of the models with $\alpha=10^{-3}\text{ and }10^{-4}$, respectively, and $\lambda=3$, colour-coded by $R_0$ (see legend). Error-bars show their 16th and 84th percentiles.}
    \label{fig:5.3}
\end{figure*}

Fig.~\ref{fig:5.3} shows the results of our comparison. Observed discs \textit{with resolved sizes} are displayed as black and grey dots for Lupus, Chamaeleon~I and Upper Sco. To avoid all the discs in each star-forming region to fall on the same position along the $x$ axis, we use the \texttt{swarmplot} function in the \texttt{python} data visualisation library \texttt{seaboarn}, that prevents them from overlapping. In addition, to give a flavour of the tentative underlying disc size distribution in each star-forming region, we draw a violin plot within the range of the observed data. Instead of showing discrete bins (e.g., histograms), \texttt{seaborn.violinplot} uses a Gaussian kernel to produce a continuous size distribution (known as kernel density estimation). The median of the data and their 16th and 84th percentiles are also indicated by the dashed and dotted lines in each violin plot. The median of the models is over-plotted using squares for $\alpha=10^{-3}$ and dots for $\alpha=10^{-4}$, with error bars displaying their 16th and 84th percentiles, obtained from the size evolution within the region age, colour-coded by their initial disc radius as in the previous plots.

As expected from the discussion so far, the theoretical predictions are very similar for all the models shown in Fig.~\ref{fig:5.3}, yet some differences can be highlighted. On the one hand, viscous models with both $\alpha=10^{-3}\text{ and }10^{-4}$ are broadly compatible with the data. The predicted sizes are somewhat at the lower end of the observed distributions, but generally within $1\sigma$ about the median of the data. On the other hand, magnetic wind models are more dependent on the $\alpha$. When $\psi=10^4$, for $\alpha=10^{-3}$ (squares) discs initially larger than $30\,{\rm au}$ are required to match the bulk of Lupus and Chamaeleon~I data. As for Upper Sco, only initial sizes of around $80\,{\rm au}$ are compatible with the median of the observed distribution. Instead, for $\alpha=10^{-4}$ the agreement improves at the age of Upper Sco, giving results similar to the viscous model ones, as expected from the reduced viscosity. Larger sizes for smaller $\alpha$ is counter-intuitive based on our previous discussion. However this happens only by the age of Upper Sco when fewer large grains are retained in models with larger $\alpha$, that are evolving (and dispersing) faster. 

In the purely magnetic wind scenario, models with $R_0<30\,{\rm au}$ are smaller than all the observed discs. However, this is not necessarily something to worry about because only resolved discs are plotted in Fig.~\ref{fig:5.3}. These correspond to about a third of the disc-bearing young stellar objects with detected (sub-)millimetre emission in each of the star-forming regions taken into account, and are those in the upper-end of the size and luminosity distribution \citep{Hendler2020}. We expect that higher-resolution observations would mitigate this bias, decreasing the median observed size, in better agreement with magnetic wind models with small $R_0$.

\paragraph*{Comments on sub-structures} None of the models can predict dust disc sizes large enough to reproduce the observed discs with $R_{\rm68,flux}\gtrsim65\,{\rm au}$. A naive solution is that of increasing $R_0$, but this would be at odds with the measured sizes of Class~0/I discs (see Section~\ref{sec:2}). More likely, discs could be larger because sub-structured.

The advent of ALMA has made it possible to detect a number of sub-structures in discs that were previously thought to be smooth (see \citealt{Andrews2020} for a recent review). The Disk Substructures at High Angular Resolution Project (DSHARP, \citealt{Andrews2018_DSHARP}) and the unbiased survey of \citet{Long2018} in Taurus supported the popular hypothesis of sub-structures being common. Nevertheless, no conclusive answer has been given in the case of fainter/smaller discs \citep{Long2019}.

Gaps and cavities, in particular, are expected to perturb the otherwise smooth and radially decreasing disc pressure profile, creating local enhancements where particles get trapped and pile up \citep{Whipple1972,Pinilla12}. The presence of sub-structures can thus affect the dust disc size evolution, because the particle radial motion is halted. However, both the extent and outcome of this process are difficult to predict because they depend on several parameters (e.g., the gap location, width, number and their ability to retain grains of different sizes for long time, see e.g., \citealt{Pinilla2020} and \citealt{Zormpas2022}). These questions will be addressed in future works.

For the time being, Fig.~\ref{fig:5.3} distinguishes between smooth discs, plotted as black dots, and sub-structured ones, shown in grey. Interestingly, lots of these have dust disc sizes much larger than those predicted by our models. However, while in Lupus we find 15 discs with sub-structures \citep{vanderMarel2018,Ansdell2018,Andrews2018_DSHARP}, only 2 were detected in Chamaeleon~I \citep{Pascucci2016} and Upper Sco \citep{Barenfeld2016,Andrews2018_DSHARP}, because of the different angular resolution of this surveys (and follow-up ones, e.g., more Lupus discs are in DSHARP, \citealt{Andrews2018_DSHARP}). Given the limited number of detected gaps in the oldest region, the extent to whom sub-structures modify our dust disc size predictions can be assessed only partially. 

\paragraph*{Longer wavelengths} Recently, \citet{Tazzari2021_obsv} published a catalogue of Lupus discs observed with ALMA in Band~3 ($3.10\,{\rm mm}$). These targets are among the brightest sources observed by \citet{Ansdell2016,Ansdell2018} in Band~6 and 7 ($1.33\,{\rm mm}$ and $0.89\,{\rm mm}$). In a companion paper, \citet{Tazzari2021_models} showed that these sources have comparable sizes in all the three bands, with the median $R_{68,{\rm flux}}$ decreasing less than 10 per cent over the explored wavelength range. 

This trend is not consistent with our smooth viscous nor magnetic disc models. In fact, we find that the 68 per cent flux radius generally follows the position in the disc where the dust grains are about the size of the observational wavelength, that \citet{Rosotti2019_radii} called the \textit{cliff radius}. For longer wavelengths the cliff radius moves inwards in the disc, because it is attained at larger grain sizes. For this reason, we expect $R_{68,{\rm flux}}$ to decrease significantly with wavelength, as opposed to what observed by \citet{Tazzari2021_models}.

This trend for $R_{68,{\rm flux}}$ being wavelength-independent is observed regardless of the presence of resolved sub-structures, suggesting that it is not due to discs being ``truncated'' by an outer disc gap (so that $R_{68,{\rm flux}}$ follows the location of the outermost gap). Thus, as suggested by \citet{Tazzari2021_models}, a possible solution comes from hypothesising that \textit{unresolved} dust traps could be ubiquitously present in the sample, explaining why large grains are retained at large radii, at odds with our models.\\

\noindent To conclude, regardless of $\psi$, models generally predict disc sizes in good agreement with those measured in Band 7, with the exception of purely magnetic wind models with a small initial radius. However, the presence of sub-structures halting the radial motion of the solids can impact our results. Our work needs to be re-assessed when enough higher-resolution data on smaller/fainter and sub-structured discs will be available, in particular at the age of Upper Sco, when the models differ the most. 

\subsection{The size-luminosity correlation}
\begin{figure*}
    \centering
    \includegraphics[width=\textwidth]{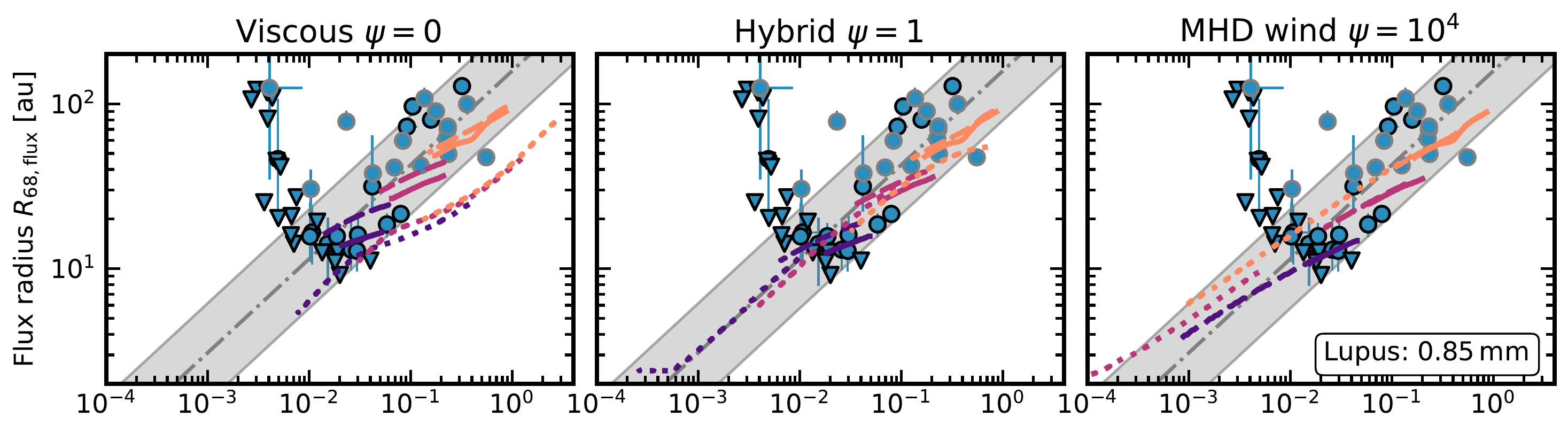}
    \includegraphics[width=\textwidth]{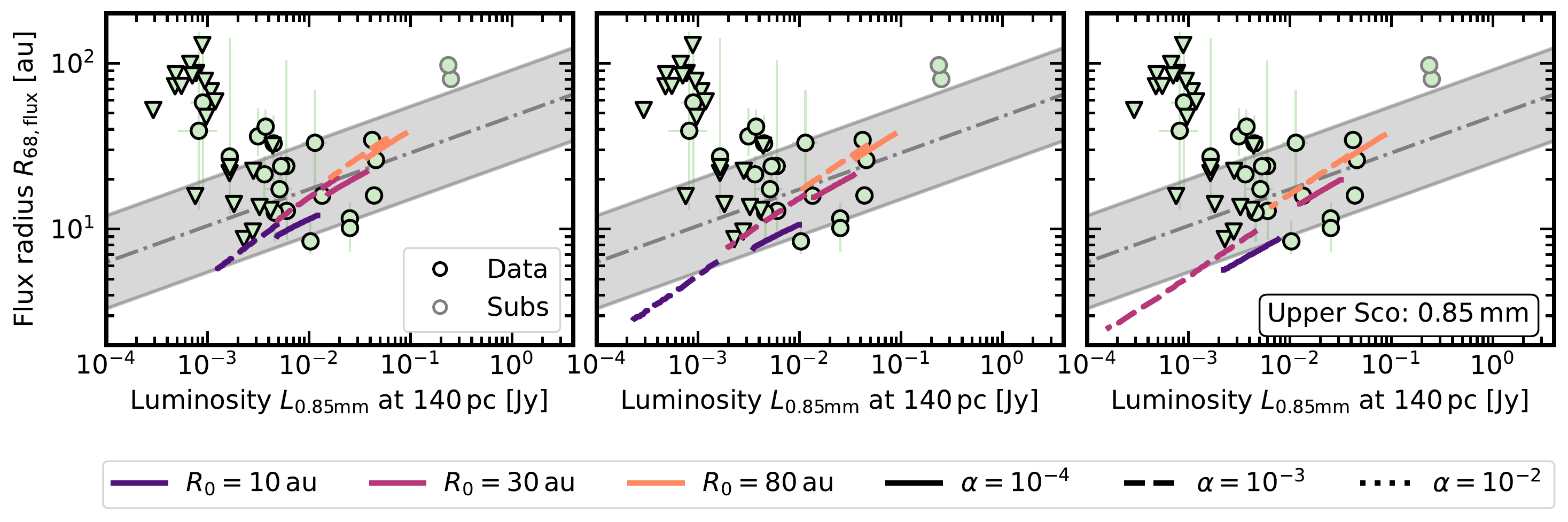}
    \caption{ALMA Band~7 size-luminosity correlation in Lupus (top row) and Upper Sco (bottom row) as a function of $\psi$. Solid, dashed and dotted lines display the evolutionary tracks for increasing values of $\alpha$, colour-coded according to $R_0$. The grey dashed-dotted lines and shaded areas, display the observed correlations and their $1\sigma$ spreads,  respectively. Data are shown as blue dots for Lupus and green dots for Upper Sco, downward-pointing triangles of the same colours are used for upper limits on the disc sizes. Grey contours are used for discs with detected sub-structures.}
    \label{fig:5.4}
\end{figure*}

\begin{figure*}
    \centering
    \includegraphics[width=\textwidth]{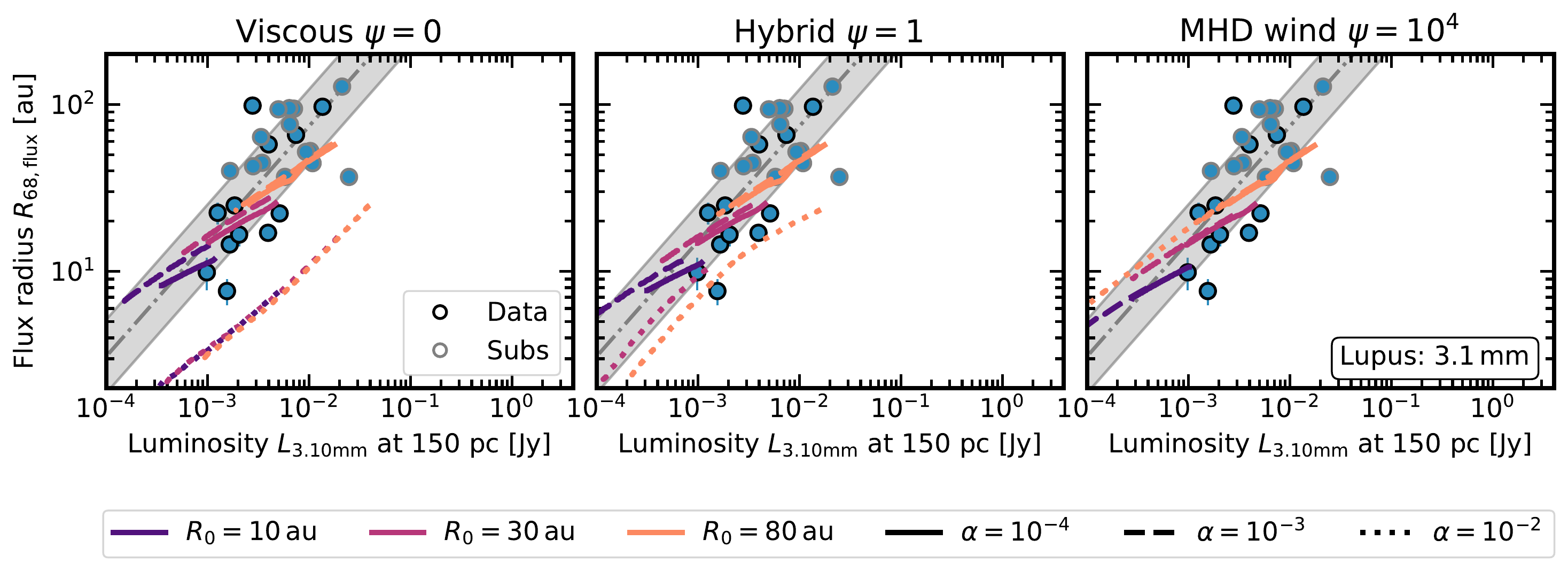}
    \caption{Same as in Fig.~\ref{fig:5.4} for ALMA Band~3 size-luminosity correlation in Lupus.}
    \label{fig:5.5}
\end{figure*}

In Fig.s~\ref{fig:5.4} and~\ref{fig:5.5} we compare models and observations in the disc dust size, $R_{\rm 68,flux}$, and luminosity, $L_{\rm mm}=F_\nu\times(d/140\,{\rm pc})^2$, plane.

A correlation between dust disc sizes and disc luminosities ($R_{\rm 68,flux}\propto L_{\rm mm}^{\alpha}$ with $\alpha\approx0.5\,{\rm to}\,0.6$) was first detected in bright Taurus discs at $0.89\,{\rm mm}$ by \citet{Tripathi2017} using SMA data. This correlation was later confirmed by \citet{Andrews2018} in a larger sample including Lupus discs and by \citet{Barenfeld2017} in Upper Sco using a different procedure. More recently, \citet{Hendler2020} studied the correlation in star-forming regions of different ages homogeneously computing dust disc sizes and luminosities. They showed that the slope of the correlation is not universal and gets flatter (and more tentative) in Upper Sco ($\alpha\approx0.22$). 

The size-luminosity correlation was originally explained as due to discs being either marginally optically thick, with an average optical depth fraction of 0.3 \citep{Tripathi2017,Andrews2018} or by the effect of dust self-scattering on fully optically thick discs \citep{Zhu2019}. Alternatively, \citet{Rosotti2019_fr} suggested that a quadratic relation between disc fluxes and sizes naturally emerges when radial drift is the main limiter of grain growth. More recently, \citet{Zormpas2022} proposed that both large sub-structured discs and small and smooth sources, where grain growth is limited by radial drift, are needed to account for the entire Lupus population. Hereafter, we explore how our models behave in the disc size \textit{vs} disc luminosity plane and how they compare with the observations. In particular, we are interested in whether a correlation exists between sizes and luminosities and how this changes with the torque ratio, the age of the region and the observational wavelength. 

In Fig.~\ref{fig:5.4} the flux radius, $R_{\rm 68,flux}$, is plotted as a function of the disc luminosity, $L_{\rm mm}$, for different values of $\psi$ after a sensitivity cut at $S_{\rm b,0.85mm}\approx6\times10^7\,{\rm Jy}\,{\rm sr^{-1}}$. Models between 1 and $3\,{\rm Myr}$ are compared with Lupus data, while models between 5 and $10\,{\rm Myr}$ are compared with Upper Sco data, in the top and bottom panels, respectively. Solid, dashed and dotted lines identify the evolutionary tracks with increasing values of $\alpha$. Observations are shown as blue dots for Lupus and green dots for Upper Sco; downward-pointing triangles of the same colours are used for disc size upper limits. The data were taken from \citet{Hendler2020} and \citet{Manara2022}. The observed correlations and their $1\sigma$ scatter are displayed as grey dashed-dotted lines and shaded areas. The best-fit values are from \citet{Hendler2020}.

Let us first focus on the top left panel where viscous models are compared with Lupus data. As shown by \citet{Rosotti2019_fr}, simulations with low $\alpha$ (solid and dashed lines) can reproduce the slope of the observed correlation  because radial drift is the main mechanism limiting grain growth. However, when $\alpha=10^{-2}$ (dotted lines), dust is fragmentation-dominated and $R_{\rm68,flux}$ does not scale as a power-law of the disc luminosity, in contrast with the bulk of the observations. Overall, our models reproduce the correlation normalisation slightly worse than those in \citet{Rosotti2019_fr}, which were a factor of two colder than ours at any given radius. In fact, several sources have luminosities $\leq0.2L_{\rm Sun}$, while our temperature profile was tailored to a Sun-like star. Furthermore, as highlighted by \citet{Zormpas2022}, also the assumed opacity model is crucial for the correlation normalisation. Some discs have larger sizes than in our models. This was already expected from Fig.~\ref{fig:5.3} and can be due to sub-structures halting radial drift. Sub-structured discs are shown in Fig.~\ref{fig:5.4} with a grey edge and partially confirm the previous hypothesis. Several sources have undetected sizes, challenging the comparison between models and data.

Similar considerations are also valid for the hybrid and purely MHD wind models. These span larger areas of the parameter space, down to $\approx10$ times lower sizes. This is particularly clear for $R_0=10\,{\rm au}$ models, as expected from Fig~\ref{fig:5.3}. An important difference with the viscous case arises when $\psi=10^4$: the slope of the size-luminosity correlation is reproduced by $\alpha=10^{-2}$ models as well. This is because in the wind-dominated models grain-growth is limited by radial drift (see the test case in Section~\ref{sec:3}) also for larger values of $\alpha$.

\citet{Hendler2020} computed homogeneously fluxes and sizes of discs in star-forming regions of different ages, showing that the slope of the size-luminosity correlation is not universal. This is not seen in our models between 5 and $10\,{\rm Myr}$, displayed in the bottom panels of Fig.~\ref{fig:5.4}. Here the evolutionary tracks have very similar slopes to those in younger models between 1 and $3\,{\rm Myr}$. This is clearly in contrast with the observed size-luminosity relation\footnote{Interestingly, purely magnetic wind models with $\lambda=1.5$ can reproduce the slope of the correlation at the age of Upper Sco, because they have similar sizes as in the case with $\lambda=3$ but much lower disc fluxes, which tilts the tracks to lower slopes.}. Nevertheless, we remark that the correlation in Upper Sco is still tentative and needs better observations to be confirmed. In particular, it is possible that several unresolved discs with large upper limits on their dust sizes contribute to flatten the correlation. Our models can reproduce some of the observed sources in the small size range, regardless of the mechanism ruling the transport of angular momentum.

\paragraph*{Longer wavelengths} \citet{Tazzari2021_models} compared sizes and luminosities at different wavelengths for a sample of Lupus discs among the brightest in the sample of \citet{Ansdell2016,Ansdell2018}. They found that a correlation is present at $0.89,\,1.33,\text{ and }3.10\,{\rm mm}$, but its normalisation and slope increase with wavelength. Similarly to Fig.~\ref{fig:5.4}, Fig.~\ref{fig:5.5} displays the size-luminosity correlation for different values of $\psi$ in ALMA Band~3. After a sensitivity cut at $S_{\rm b,3.10mm}\approx1\times10^7\,{\rm Jy}\,{\rm sr^{-1}}$ (corresponding to an integration time of $2\,{\rm min}$ and an angular resolution of $0.35\,{\rm arcsec}$ as in the configuration used for the 18 brightest sources in the sample of \citealt{Tazzari2021_obsv}), our models between 1 and $3\,{\rm Myr}$ are compared with Lupus data \citep{Tazzari2021_obsv,Tazzari2021_models}.

As can be seen, the slope of the models is very similar to the one in Fig.~\ref{fig:5.4}. Indeed, under the assumption that the flux radius correlation is driven by the grains being in the drift regime, \citet{Rosotti2019_fr} showed that changing the observational wavelength only affects the correlation normalisation that is expected to increase proportionally to the wavelength squared. However, in the data, the correlation slope is slightly different\footnote{Recently \citet{Zormpas2022} published a more comprehensive analysis of the size-luminosity correlation in Lupus, where they suggested that smooth viscous discs are not consistent with the observed Band~3 correlation. However, \citet{Zormpas2022} compared their results with the correlation coefficients obtained for face-on discs (see \citealt{Tazzari2021_models}). Instead, we use the integrated flux for consistency with \citet{Hendler2020}. Interestingly, the largest difference between the correlation slope is obtained in Band~3, where discs are more optically thin than in Band~7 and the effect of the disc inclination should be less important.} between the two bands: $\approx0.57\text{ to }0.61$ in Band~7 \citep{Hendler2020,Tazzari2021_models} and $\approx0.69$ in Band~3 \citep{Tazzari2021_models}, potentially because of the absence of faint discs in the sample of \citet{Tazzari2021_obsv}. Most importantly, our models are smaller than roughly half of the sources in \citep{Tazzari2021_models}, several of which show evidence of sub-structures, as already highlighted in the previous sub-section.\\ 

\noindent To sum up, viscous and magnetic wind models behave similarly in the size-luminosity plane and can explain the observed correlation when they are in the drift-dominated regime. However, the non-universality of the correlation slope in the data and the presence of discs too large for our models suggest that the correlation is not only/primarily shaped by the process controlling grain-growth (e.g., radial-drift). Sub-structures trapping solids efficiently could be a possible explanation for these discrepancies as is discussed in \citet{Zormpas2022}, both in the viscous and MHD-wind case.

\section{Discussion}\label{sec:6}

\subsection{Grain growth in the absence of turbulence} Our predictions on the secular evolution of dust disc sizes depend on the underlying grain growth model. In fact, the implementation of a full coagulation routine is computationally expensive and prevents the exploration of a large parameter space. For this reason we adopted the simplified treatment of grain growth provided by \citet{Birnstiel2012}, which reproduces the results of more complex calculation but is less time-consuming. This model was benchmarked against a grid of 39 full simulations with $10^{-5}\leq\alpha_{\rm SS}\leq10^{-3}$. Such a range is consistent with the parameter space we explored in the case of viscous and hybrid models, but not in the purely magnetic wind scenario, where $10^{-8}\leq\alpha_{\rm SS}\leq10^{-6}$. What is more, \citet{Birnstiel2012} considered turbulence as the main source of particle relative velocities, and hence the driver of grain growth. However, this is not true at the low viscosities considered in our magnetic wind models. 

It can be hypothesised that grain growth is inhibited in laminar discs: when only Brownian motion and vertical settling are considered as a source of dust relative velocities, $a_{\rm max}\approx1\,{\rm cm}$ at $1\,{\rm au}$ after $1\,{\rm Myr}$ \citep{Safronov1969,Dullemond&Dominik2005}. Instead, if vertical mixing and turbulent relative velocities are included, $a_{\rm max}\approx10^4\,{\rm to}\,10^5\,{\rm cm}$ \citep{Dullemond&Dominik2005,Brauer2008}. Nevertheless, \citet{Brauer2008} showed that when radial drift and fragmentation are also taken into account, this picture changes. In the turbulent case the maximum size grains can grow to is dramatically reduced because of destructive collisions between very fast grains, while in the laminar case the differential radial motion becomes an important source of relative particle velocities that allows for coagulation. In the models of \citet{Brauer2008}, after $10^4\,{\rm yr}$ solids are 20 times larger when $\alpha_{\rm SS}=10^{-10}$ than for $\alpha_{\rm SS}=10^{-3}$ (see Fig.~14 in \citealt{Brauer2008}), showing that grains can grow also in laminar discs and potentially to larger sizes than in turbulent ones. 

Provided that solids can grow in the absence of turbulence, we are confident in the two population predictions in the laminar case. Even though the growth time scale was not tailored to reproduce the purely magnetic wind scenario, it only influences our results in the very first stages of disc evolution ($t\lesssim0.1\,{\rm Myr}$), then radial drift becomes the dominant process setting the dust disc sizes. 

In the absence of strong turbulence, relative particle velocities are determined by differential radial drift, a mechanism potentially leading to particle fragmentation. \citet{Birnstiel2012} showed that fragmentation by differential drift can be safely overlooked in the outer disc, where the radial drift limit becomes dominant early on. However, when $\alpha_{\rm SS}$ is very low, it can be important in the inner disc. For this reason, fragmentation by differential drift is implemented by default in our code. All in all, it sets the maximum grain size only in the $R\lesssim1\,{\rm au}$ region of the disc when $t\lesssim1\,{\rm Myr}$, as long as the dust-to-gas ratio is large, $\epsilon\gtrsim 0.0013\times R^{0.5}$. We conclude, as \citet{Birnstiel2012} did, that this fragmentation mechanism is generally unimportant (but see \citealt{Pinilla2021} in the case of lower fragmentation velocities).

\subsection{Dust entrainment in magnetic disc winds} 
MHD disc winds are expected to be strong enough to uplift small dust grains as supported by the evidence of solids in disc outflows \citep[e.g.,][]{Bans2012,Ellerbroek2014}. The entrainment and thermal processing of dust in magnetic winds has also been proposed as an explanation for the presence of crystalline silicates in the outer disc, where temperatures are too low for their \textit{in situ} formation through thermal annealing \citep{Salmeron&Ireland2012,Giacalone2019}.

In our models, however, we neglect dust removal in the wind. This is a reasonable approximation as long as dust particles can grow rapidly to scales that are sufficiently larger than the maximum grain size that can be entrained in the wind, $a_{\rm crit}$. In what follows, we discuss values of $a_{\rm crit}$ from recent literature studies and estimate when the dust mass loss in the wind can be significant for our results.

\citet{Miyake2016} first studied dust vertical motion in the presence of magnetic winds, but adopted a fixed background for gas, resulting from previous 3D \textit{local} shearing box simulations. \citet{Giacalone2019}, instead, employed \textit{global} steady-state semi-analytical solutions for the gas and also considered dust radial motion. They showed that only dust particles with radius smaller than $a_{\rm crit}\approx 0.1\,{\rm to}\,1\,\mu{\rm m}$ can be uplifted by the wind, with $\approx30\%$ of them re-entering the disc at larger radii. Similar maximum entrained particle sizes were inferred by \citet{Rodenkirch&Dullemond2022}, who performed 2D non-ideal MHD simulations employing different dust species. 

Nevertheless, $a_{\rm crit}$ scales linearly with the \textit{gas} mass loss rate in the wind. \citet{Giacalone2019} chose $\dot{M}_{\rm g,w}=3.5\times10^{-8}\,M_{\rm Sun}\,{\rm yr}^{-1}$, corresponding to an implausibly high efficiency wind with $\lambda=108.4$, and similar values were inferred by \citet{Rodenkirch&Dullemond2022} in discs with warm (i.e. with thermal contributions) and cold (i.e. purely magnetic) winds with $10^4\leq\beta_0\leq10^5$. This suggests that larger grains can be uplifted when the mass loss rates in the wind are larger, for example in early disc evolutionary phases (as see in \citealt{Miotello2014} and suggested by \citealt{Wong2016}).

More recently, \citet{Booth&Clarke2021} modelled the entrainment of dust in ionised winds, showing that it is prompted by the delivery of small grains to the wind base, which is induced by advection in the wind rather than turbulent diffusion. We use the model of \citet{Booth&Clarke2021} to infer the maximum size of grains delivered to the wind in our simulations and estimating the total \textit{dust} mass loss rate in the wind, $\dot{M}_{\rm d,w}$, as:
\begin{equation}\label{eq:6}
    \dot{M}_{\rm d,w}=\int_{R_{\rm in}}^{R_{\rm out}}f\dot{\Sigma}_{\rm g,w}\epsilon2\pi RdR.
\end{equation}
Here $\dot{\Sigma}_{\rm g,w}$ is the \textit{gas} mass loss rate in the wind computed as in Eq.~\ref{eq:2}, $\epsilon$ is the dust-to-gas ratio and $f$ is the fraction of grains entrained in the wind. To compute such fraction we assume that the number of grains of sizes between $a$ and $a+da$ follows a power-law distribution of the grain size, $n(a)da\propto a^{-q}da$, leading to:
\begin{equation}\label{eq:7}
    f\approx\min\left[1,\left(\dfrac{a_{\rm crit}}{a_{\rm max}}\right)^{4-q}\right].
\end{equation}
In this work we assume a MRN distribution \citep{Mathis1977}, with $q=3.5$, which is appropriate for a collisional distribution of grains, such as that of small growing dust particles. Nevertheless, using more top-heavy grain size distributions (up to $q=2.5$), more suitable for dust grains sizes dominated by radial drift \citep{Birnstiel2012}, our results remain similar.

Using Eq.~\ref{eq:6} we can estimate the \textit{dust} mass lost in the wind summing over each simulation time step. Generally, less than 1 to 2 per cent of the initial dust mass is removed by the wind. Notable exceptions are initially small discs with very strong winds ($R_0=10\,{\rm au}$, $\alpha\gtrsim10^{-2}$, $\lambda=1.5$), where this fraction can reach 15 to 40 per cent. In such cases, however, a comparison between dust disc sizes in viscous and magnetic wind models on secular time scales was already prohibitive, because of the very fast dispersal of dust discussed in Sections~\ref{sec:4} and~\ref{sec:5}, in the MHD-wind case. 

Using the $a_{\rm crit}$ of \citet{Giacalone2019} almost identical results are obtained. Our inferences refer to the first $3\,{\rm Myr}$ of disc lifetime but are not expected to change over longer times as the bulk of the grains is removed in the initial time step ($t\lesssim2.5\times10^4\,{\rm yr}$), where $a_{\rm crit}>a_{\rm max}$ throughout the disc. Finally, it must be mentioned that the grain growth time scale is crucial for the determination of the dust mass loss rates in the wind. If grains can grow faster in MHD discs than predicted by the two-population model, our previous estimates are upper limits.

To sum up, we are confident that dust entrainment in MHD winds can be overlooked in those models with $\alpha\approx10^{-3}\,{\rm to}\,10^{-4}$ we used to compare disc sizes between viscous and MHD evolving discs.


\subsection{Late time dispersal} Discs do not simply fade away with time but are expected to disperse abruptly \citep[e.g.,][]{Fedele2010}. In the traditional framework of viscous evolution, photo-evaporation has been successfully proposed as an efficient dispersal mechanism because of its possibility to account for the observed inside-out disc clearing \citep{Koepferl&Ercolano2013}. On top of viscous evolution, internal photo-evaporation \citep{Clarke2001,Owen2010} brings about a phase of so-called ``photo-evaporation starved accretion" \citep{Drake2009}. The disc becomes progressively depleted in the region of maximum wind mass loss rate (where $R\approx GM_*/c_\mathrm{s}^2$, roughly at tens of au), then a gap opens further in. The inner disc is viscously drained on the (smaller) gap edge time scale, while the outer disc is progressively eroded by the wind. We expect photo-evaporation to impact disc size evolution as discussed before in the case of gaps, because the wind-opened cavity can retain dust, forming bright rings.

\begin{figure}
    \centering
    \includegraphics[width=0.85\columnwidth]{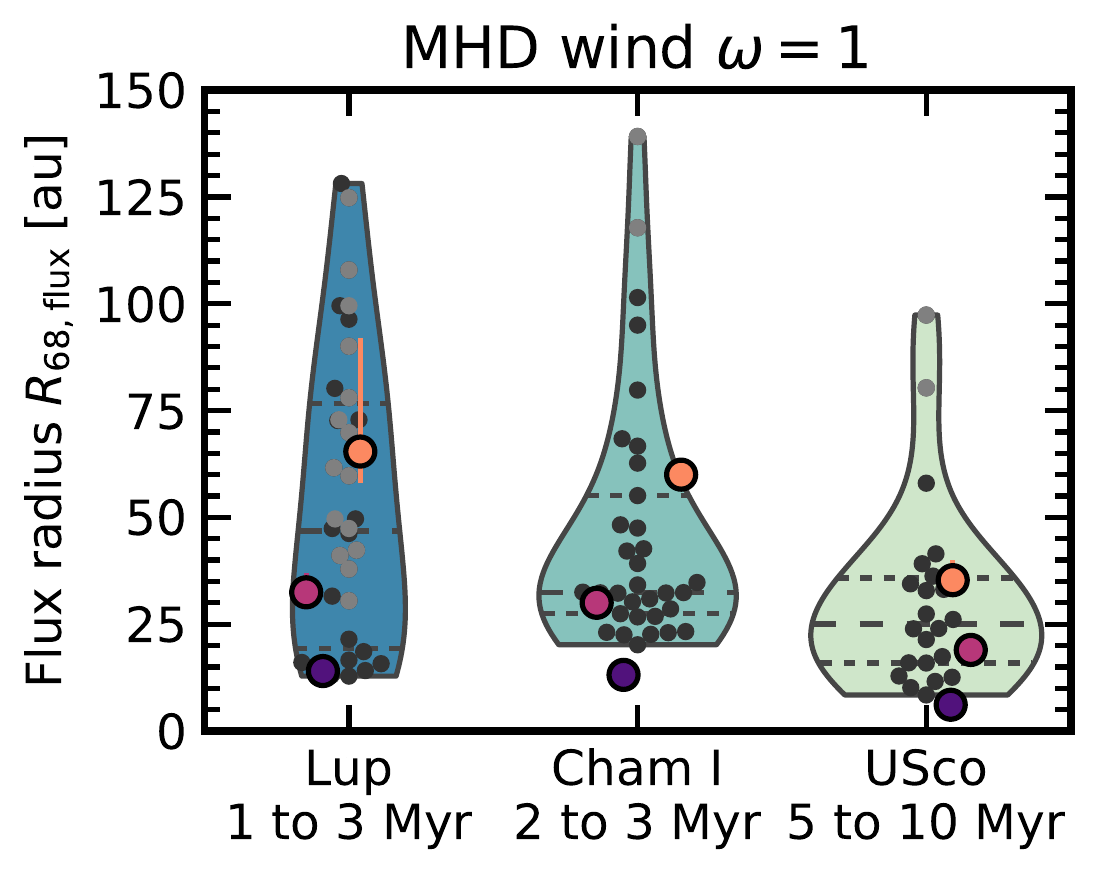}
    \caption{Same as in Fig.~\ref{fig:5.3} for $\omega=1$ models with $\alpha_{\rm DW}(t=0)=10^{-4}$ and $\lambda=3$.}
    \label{fig:6.1}
\end{figure}

On the contrary, discs evolving under the effect of MHD winds naturally disperse if a dependence of the angular momentum transport coefficient on the disc mass, $\alpha_{\rm DW}\propto M_{\rm disc}^{-\omega}$ with $\omega>0$, is considered \citep{Tabone2022a}. These models however, cannot reproduce the inside-out clearing predicted by photometric disc surveys\footnote{Some models, such as those of \citet{Suzuki2016}, predict a inner depletion of the disc, but this outcome depends on the prescription for the wind/disc ionisation and is not naturally arising as in the case of photo-evaporation.}. Mass and flux sizes when $\omega=1$ are plotted in Appendix~\ref{app:2} and show very similar results as the purely magnetic wind case with $\omega=0$ considered so far. In Fig.~\ref{fig:6.1} we test how the dust disc sizes of these models compare with the observations as in Fig.~\ref{fig:5.3}. However, here we only consider disc models with $\alpha=10^{-4}$ at $t=0$. This is because even for small discs with $R_0=10\,{\rm au}$, their dispersion time scale, $t_{\rm disp}\lesssim10\,{\rm Myr}$, is long enough for a comparison with Upper Sco discs. As is clear from the plot, $\omega=1$ models behave as the purely magnetic wind ones (with $\omega=0$) in Fig.~\ref{fig:5.3}: they reproduce well the range of observed sizes within $1\sigma$ about the median. Again the largest discs in \citet{Hendler2020} cannot be accounted for by smooth models.

\section{Conclusions}\label{sec:7}
This paper is devoted to making predictions on dust disc sizes and comparing them with the currently available observations in nearby star-forming regions. To do so, we ran a number of one-dimensional gas and dust models with the aim of exploring a large region of the parameter space, varying the angular momentum transport efficiency, $\alpha$, the torque ratio, $\psi$, the initial scale radius, $R_0$, the lever arm parameter, $\lambda$, and the dependence of $\alpha$ with radius in the purely magnetic wind case. The results of our exercise are summarised hereafter.
\begin{itemize}
    \item The mass radius and the 95 per cent flux radius have different behaviours in the viscous ($\psi=0$) and magnetic wind ($\psi=10^4$) case, expanding in the former and shrinking or plateauing in the latter (when $\alpha\gtrsim10^{-3}$);
    \item Current observations cannot discriminate between viscous and magnetic wind models because they are not sensitive enough to recover the potentially faint and viscously expanding outer disc regions. Deeper surveys in star-forming regions with a large age difference could be fruitful to observe the predicted differences between dust disc size evolution in viscous and magnetic wind models. Nonetheless, such observations could be challenging even for ALMA;
    \item The model predictions are in agreement with the currently available observationally-inferred dust disc sizes from Lupus, Chamaeleon~I and Upper Sco Band~7 ($0.89\,{\rm mm}$) data within $1\sigma$ about the median, regardless of $\psi$. Some purely magnetic wind models predict smaller sizes than in the data. This is not necessarily a problem given the limited angular resolution of the current surveys. None of the models can explain very large discs. A possible explanation is that several of these sources are sub-structured;
    \item Our models predict smaller flux sizes for longer wavelengths because radial drift segregates the largest grains in the inner disc. This is in contrast with disc sizes inferred from Band~3 ($3.10\,{\rm mm}$) Lupus observations, that are in line with Band~7 ones;
    \item In Lupus, our models predict a similar size-luminosity correlation as the one detected at $0.89\,{\rm mm}$ and (marginally) at $3.10\,{\rm mm}$. In Upper Sco, however, the slope of the correlation is steeper in the models than in the data. 
\end{itemize}
A more fruitful comparison with the data would be possible exploring the model dependence on other key parameters, such as the initial disc mass, stellar mass, temperature... and ultimately running disc population synthesis models where the initial conditions are randomly chosen from observationally-motivated parameter distributions. Predictions on the size evolution in the presence of sub-structures is also required to interpret the data.

\section*{Acknowledgements}
We thank the anonymous reviewer for their comments that helped to improve the manuscript. FZ acknowledges support from STFC and Cambridge Trust for a Ph.D. studentship. GR acknowledges support from the Netherlands Organisation for Scientific Research (NWO, program number 016.Veni.192.233) and from an STFC Ernest Rutherford Fellowship (grant number ST/T003855/1). This work was supported by the European Union’s Horizon 2020 research and innovation programme under the Marie Sklodowska Curie grant agreement number 823823 (DUSTBUSTERS).

Software: \texttt{numpy} \citep{numpy20_2020Natur.585..357H}, \texttt{matplotlib} \citep{matplotlib_Hunter:2007}, \texttt{scipy} \citep{scipy_2020SciPy-NMeth}, \texttt{JupyterNotebook} \citep{Jupyter_nootbok}, \texttt{seaborn} \citet{Waskom2021}.


\section*{Data Availability}
The code used in this paper is publicly available on GitHub at
\texttt{github.com/rbooth200/DiscEvolution}. The data underlying
this paper are available in the ALMA archive as explained in \citet{Hendler2020} and \citet{Tazzari2021_models,Tazzari2021_obsv}. The PPVII chapter summary table of \citet{Manara2022} can be found at \texttt{ppvii.org/chapter/15/}.



\bibliographystyle{mnras}
\bibliography{refs}

\begin{thebibliography}{}
\makeatletter
\relax
\def\mn@urlcharsother{\let\do\@makeother \do\$\do\&\do\#\do\^\do\_\do\%\do\~}
\def\mn@doi{\begingroup\mn@urlcharsother \@ifnextchar [ {\mn@doi@}
  {\mn@doi@[]}}
\def\mn@doi@[#1]#2{\def\@tempa{#1}\ifx\@tempa\@empty \href
  {http://dx.doi.org/#2} {doi:#2}\else \href {http://dx.doi.org/#2} {#1}\fi
  \endgroup}
\def\mn@eprint#1#2{\mn@eprint@#1:#2::\@nil}
\def\mn@eprint@arXiv#1{\href {http://arxiv.org/abs/#1} {{\tt arXiv:#1}}}
\def\mn@eprint@dblp#1{\href {http://dblp.uni-trier.de/rec/bibtex/#1.xml}
  {dblp:#1}}
\def\mn@eprint@#1:#2:#3:#4\@nil{\def\@tempa {#1}\def\@tempb {#2}\def\@tempc
  {#3}\ifx \@tempc \@empty \let \@tempc \@tempb \let \@tempb \@tempa \fi \ifx
  \@tempb \@empty \def\@tempb {arXiv}\fi \@ifundefined
  {mn@eprint@\@tempb}{\@tempb:\@tempc}{\expandafter \expandafter \csname
  mn@eprint@\@tempb\endcsname \expandafter{\@tempc}}}

\bibitem[\protect\citeauthoryear{{Andrews}}{{Andrews}}{2020}]{Andrews2020}
{Andrews} S.~M.,  2020, \mn@doi [\araa] {10.1146/annurev-astro-031220-010302},
  \href {https://ui.adsabs.harvard.edu/abs/2020ARA&A..58..483A} {58, 483}

\bibitem[\protect\citeauthoryear{{Andrews}, {Terrell}, {Tripathi}, {Ansdell},
  {Williams}  \& {Wilner}}{{Andrews} et~al.}{2018a}]{Andrews2018}
{Andrews} S.~M.,  {Terrell} M.,  {Tripathi} A.,  {Ansdell} M.,  {Williams}
  J.~P.,   {Wilner} D.~J.,  2018a, \mn@doi [\apj] {10.3847/1538-4357/aadd9f},
  \href {https://ui.adsabs.harvard.edu/abs/2018ApJ...865..157A} {865, 157}

\bibitem[\protect\citeauthoryear{{Andrews} et~al.,}{{Andrews}
  et~al.}{2018b}]{Andrews2018_DSHARP}
{Andrews} S.~M.,  et~al., 2018b, \mn@doi [\apjl] {10.3847/2041-8213/aaf741},
  \href {https://ui.adsabs.harvard.edu/abs/2018ApJ...869L..41A} {869, L41}

\bibitem[\protect\citeauthoryear{{Ansdell} et~al.,}{{Ansdell}
  et~al.}{2016}]{Ansdell2016}
{Ansdell} M.,  et~al., 2016, \mn@doi [\apj] {10.3847/0004-637X/828/1/46}, \href
  {https://ui.adsabs.harvard.edu/abs/2016ApJ...828...46A} {828, 46}

\bibitem[\protect\citeauthoryear{{Ansdell} et~al.,}{{Ansdell}
  et~al.}{2018}]{Ansdell2018}
{Ansdell} M.,  et~al., 2018, \mn@doi [\apj] {10.3847/1538-4357/aab890}, \href
  {https://ui.adsabs.harvard.edu/abs/2018ApJ...859...21A} {859, 21}

\bibitem[\protect\citeauthoryear{{Arakawa}, {Matsumoto}  \& {Honda}}{{Arakawa}
  et~al.}{2021}]{Arakawa2021}
{Arakawa} S.,  {Matsumoto} Y.,   {Honda} M.,  2021, \mn@doi [\apj]
  {10.3847/1538-4357/ac157e}, \href
  {https://ui.adsabs.harvard.edu/abs/2021ApJ...920...27A} {920, 27}

\bibitem[\protect\citeauthoryear{{Armitage}, {Simon}  \& {Martin}}{{Armitage}
  et~al.}{2013}]{Armitage2013}
{Armitage} P.~J.,  {Simon} J.~B.,   {Martin} R.~G.,  2013, \mn@doi [\apjl]
  {10.1088/2041-8205/778/1/L14}, \href
  {https://ui.adsabs.harvard.edu/abs/2013ApJ...778L..14A} {778, L14}

\bibitem[\protect\citeauthoryear{{Bai}}{{Bai}}{2013}]{Bai2013}
{Bai} X.-N.,  2013, \mn@doi [\apj] {10.1088/0004-637X/772/2/96}, \href
  {https://ui.adsabs.harvard.edu/abs/2013ApJ...772...96B} {772, 96}

\bibitem[\protect\citeauthoryear{{Bai}}{{Bai}}{2016}]{Bai2016}
{Bai} X.-N.,  2016, \mn@doi [\apj] {10.3847/0004-637X/821/2/80}, \href
  {https://ui.adsabs.harvard.edu/abs/2016ApJ...821...80B} {821, 80}

\bibitem[\protect\citeauthoryear{{Bai}}{{Bai}}{2017}]{Bai2017}
{Bai} X.-N.,  2017, \mn@doi [\apj] {10.3847/1538-4357/aa7dda}, \href
  {https://ui.adsabs.harvard.edu/abs/2017ApJ...845...75B} {845, 75}

\bibitem[\protect\citeauthoryear{{Bai} \& {Stone}}{{Bai} \&
  {Stone}}{2013}]{Bai&Stone2013}
{Bai} X.-N.,  {Stone} J.~M.,  2013, \mn@doi [\apj]
  {10.1088/0004-637X/769/1/76}, \href
  {https://ui.adsabs.harvard.edu/abs/2013ApJ...769...76B} {769, 76}

\bibitem[\protect\citeauthoryear{{Balbus} \& {Hawley}}{{Balbus} \&
  {Hawley}}{1991}]{Balbus&Hawley1991}
{Balbus} S.~A.,  {Hawley} J.~F.,  1991, \mn@doi [\apj] {10.1086/170270}, \href
  {https://ui.adsabs.harvard.edu/abs/1991ApJ...376..214B} {376, 214}

\bibitem[\protect\citeauthoryear{{Bans} \& {K{\"o}nigl}}{{Bans} \&
  {K{\"o}nigl}}{2012}]{Bans2012}
{Bans} A.,  {K{\"o}nigl} A.,  2012, \mn@doi [\apj]
  {10.1088/0004-637X/758/2/100}, \href
  {https://ui.adsabs.harvard.edu/abs/2012ApJ...758..100B} {758, 100}

\bibitem[\protect\citeauthoryear{{Banzatti}, {Pascucci}, {Edwards}, {Fang},
  {Gorti}  \& {Flock}}{{Banzatti} et~al.}{2019}]{Banzatti2019}
{Banzatti} A.,  {Pascucci} I.,  {Edwards} S.,  {Fang} M.,  {Gorti} U.,
  {Flock} M.,  2019, \mn@doi [\apj] {10.3847/1538-4357/aaf1aa}, \href
  {https://ui.adsabs.harvard.edu/abs/2019ApJ...870...76B} {870, 76}

\bibitem[\protect\citeauthoryear{{Barenfeld}, {Carpenter}, {Ricci}  \&
  {Isella}}{{Barenfeld} et~al.}{2016}]{Barenfeld2016}
{Barenfeld} S.~A.,  {Carpenter} J.~M.,  {Ricci} L.,   {Isella} A.,  2016,
  \mn@doi [\apj] {10.3847/0004-637X/827/2/142}, \href
  {https://ui.adsabs.harvard.edu/abs/2016ApJ...827..142B} {827, 142}

\bibitem[\protect\citeauthoryear{{Barenfeld}, {Carpenter}, {Sargent}, {Isella}
  \& {Ricci}}{{Barenfeld} et~al.}{2017}]{Barenfeld2017}
{Barenfeld} S.~A.,  {Carpenter} J.~M.,  {Sargent} A.~I.,  {Isella} A.,
  {Ricci} L.,  2017, \mn@doi [\apj] {10.3847/1538-4357/aa989d}, \href
  {https://ui.adsabs.harvard.edu/abs/2017ApJ...851...85B} {851, 85}

\bibitem[\protect\citeauthoryear{{B{\'e}thune}, {Lesur}  \&
  {Ferreira}}{{B{\'e}thune} et~al.}{2017}]{Bethune2017}
{B{\'e}thune} W.,  {Lesur} G.,   {Ferreira} J.,  2017, \mn@doi [\aap]
  {10.1051/0004-6361/201630056}, \href
  {https://ui.adsabs.harvard.edu/abs/2017A&A...600A..75B} {600, A75}

\bibitem[\protect\citeauthoryear{{Birnstiel} \& {Andrews}}{{Birnstiel} \&
  {Andrews}}{2014}]{Birnstiel&Andrews2014}
{Birnstiel} T.,  {Andrews} S.~M.,  2014, \mn@doi [\apj]
  {10.1088/0004-637X/780/2/153}, \href
  {https://ui.adsabs.harvard.edu/abs/2014ApJ...780..153B} {780, 153}

\bibitem[\protect\citeauthoryear{{Birnstiel}, {Dullemond}  \&
  {Brauer}}{{Birnstiel} et~al.}{2010}]{Birnstiel2010}
{Birnstiel} T.,  {Dullemond} C.~P.,   {Brauer} F.,  2010, \mn@doi [\aap]
  {10.1051/0004-6361/200913731}, \href
  {https://ui.adsabs.harvard.edu/abs/2010A&A...513A..79B} {513, A79}

\bibitem[\protect\citeauthoryear{{Birnstiel}, {Klahr}  \&
  {Ercolano}}{{Birnstiel} et~al.}{2012}]{Birnstiel2012}
{Birnstiel} T.,  {Klahr} H.,   {Ercolano} B.,  2012, \mn@doi [\aap]
  {10.1051/0004-6361/201118136}, \href
  {https://ui.adsabs.harvard.edu/abs/2012A&A...539A.148B} {539, A148}

\bibitem[\protect\citeauthoryear{{Birnstiel}, {Fang}  \&
  {Johansen}}{{Birnstiel} et~al.}{2016}]{Birnstiel2016}
{Birnstiel} T.,  {Fang} M.,   {Johansen} A.,  2016, \mn@doi [\ssr]
  {10.1007/s11214-016-0256-1}, \href
  {https://ui.adsabs.harvard.edu/abs/2016SSRv..205...41B} {205, 41}

\bibitem[\protect\citeauthoryear{{Blandford} \& {Payne}}{{Blandford} \&
  {Payne}}{1982}]{Blandford&Payne1982}
{Blandford} R.~D.,  {Payne} D.~G.,  1982, \mn@doi [\mnras]
  {10.1093/mnras/199.4.883}, \href
  {https://ui.adsabs.harvard.edu/abs/1982MNRAS.199..883B} {199, 883}

\bibitem[\protect\citeauthoryear{{Bohlin}, {Savage}  \& {Drake}}{{Bohlin}
  et~al.}{1978}]{Bohlin1978}
{Bohlin} R.~C.,  {Savage} B.~D.,   {Drake} J.~F.,  1978, \mn@doi [\apj]
  {10.1086/156357}, \href
  {https://ui.adsabs.harvard.edu/abs/1978ApJ...224..132B} {224, 132}

\bibitem[\protect\citeauthoryear{{Booth} \& {Clarke}}{{Booth} \&
  {Clarke}}{2021}]{Booth&Clarke2021}
{Booth} R.~A.,  {Clarke} C.~J.,  2021, \mn@doi [\mnras]
  {10.1093/mnras/stab090}, \href
  {https://ui.adsabs.harvard.edu/abs/2021MNRAS.502.1569B} {502, 1569}

\bibitem[\protect\citeauthoryear{{Booth} \& {Owen}}{{Booth} \&
  {Owen}}{2020}]{Booth&Owen2020}
{Booth} R.~A.,  {Owen} J.~E.,  2020, \mn@doi [\mnras] {10.1093/mnras/staa578},
  \href {https://ui.adsabs.harvard.edu/abs/2020MNRAS.493.5079B} {493, 5079}

\bibitem[\protect\citeauthoryear{{Booth}, {Clarke}, {Madhusudhan}  \&
  {Ilee}}{{Booth} et~al.}{2017}]{Booth2017}
{Booth} R.~A.,  {Clarke} C.~J.,  {Madhusudhan} N.,   {Ilee} J.~D.,  2017,
  \mn@doi [\mnras] {10.1093/mnras/stx1103}, \href
  {https://ui.adsabs.harvard.edu/abs/2017MNRAS.469.3994B} {469, 3994}

\bibitem[\protect\citeauthoryear{{Booth} et~al.,}{{Booth}
  et~al.}{2021}]{Booth2021}
{Booth} A.~S.,  et~al., 2021, \mn@doi [\apjs] {10.3847/1538-4365/ac1ad4}, \href
  {https://ui.adsabs.harvard.edu/abs/2021ApJS..257...16B} {257, 16}

\bibitem[\protect\citeauthoryear{{Brauer}, {Dullemond}  \& {Henning}}{{Brauer}
  et~al.}{2008}]{Brauer2008}
{Brauer} F.,  {Dullemond} C.~P.,   {Henning} T.,  2008, \mn@doi [\aap]
  {10.1051/0004-6361:20077759}, \href
  {https://ui.adsabs.harvard.edu/abs/2008A&A...480..859B} {480, 859}

\bibitem[\protect\citeauthoryear{{Chiang} \& {Goldreich}}{{Chiang} \&
  {Goldreich}}{1997}]{Chiang&Goldreich1997}
{Chiang} E.~I.,  {Goldreich} P.,  1997, \mn@doi [\apj] {10.1086/304869}, \href
  {https://ui.adsabs.harvard.edu/abs/1997ApJ...490..368C} {490, 368}

\bibitem[\protect\citeauthoryear{{Clarke}, {Gendrin}  \& {Sotomayor}}{{Clarke}
  et~al.}{2001}]{Clarke2001}
{Clarke} C.~J.,  {Gendrin} A.,   {Sotomayor} M.,  2001, \mn@doi [\mnras]
  {10.1046/j.1365-8711.2001.04891.x}, \href
  {https://ui.adsabs.harvard.edu/abs/2001MNRAS.328..485C} {328, 485}

\bibitem[\protect\citeauthoryear{{Comer{\'o}n}}{{Comer{\'o}n}}{2008}]{Comeron2008}
{Comer{\'o}n} F.,  2008, {The Lupus Clouds}.
p.~295

\bibitem[\protect\citeauthoryear{{Dipierro}, {Laibe}, {Alexander}  \&
  {Hutchison}}{{Dipierro} et~al.}{2018}]{Dipierro2018}
{Dipierro} G.,  {Laibe} G.,  {Alexander} R.,   {Hutchison} M.,  2018, \mn@doi
  [\mnras] {10.1093/mnras/sty1701}, \href
  {https://ui.adsabs.harvard.edu/abs/2018MNRAS.479.4187D} {479, 4187}

\bibitem[\protect\citeauthoryear{{Drake}, {Ercolano}, {Flaccomio}  \&
  {Micela}}{{Drake} et~al.}{2009}]{Drake2009}
{Drake} J.~J.,  {Ercolano} B.,  {Flaccomio} E.,   {Micela} G.,  2009, \mn@doi
  [\apjl] {10.1088/0004-637X/699/1/L35}, \href
  {https://ui.adsabs.harvard.edu/abs/2009ApJ...699L..35D} {699, L35}

\bibitem[\protect\citeauthoryear{{Dullemond} \& {Dominik}}{{Dullemond} \&
  {Dominik}}{2005}]{Dullemond&Dominik2005}
{Dullemond} C.~P.,  {Dominik} C.,  2005, \mn@doi [\aap]
  {10.1051/0004-6361:20042080}, \href
  {https://ui.adsabs.harvard.edu/abs/2005A&A...434..971D} {434, 971}

\bibitem[\protect\citeauthoryear{{Ellerbroek} et~al.,}{{Ellerbroek}
  et~al.}{2014}]{Ellerbroek2014}
{Ellerbroek} L.~E.,  et~al., 2014, \mn@doi [\aap]
  {10.1051/0004-6361/201323092}, \href
  {https://ui.adsabs.harvard.edu/abs/2014A&A...563A..87E} {563, A87}

\bibitem[\protect\citeauthoryear{{Facchini}, {Birnstiel}, {Bruderer}  \& {van
  Dishoeck}}{{Facchini} et~al.}{2017}]{Facchini2017}
{Facchini} S.,  {Birnstiel} T.,  {Bruderer} S.,   {van Dishoeck} E.~F.,  2017,
  \mn@doi [\aap] {10.1051/0004-6361/201630329}, \href
  {https://ui.adsabs.harvard.edu/abs/2017A&A...605A..16F} {605, A16}

\bibitem[\protect\citeauthoryear{{Facchini} et~al.,}{{Facchini}
  et~al.}{2019}]{Facchini2019}
{Facchini} S.,  et~al., 2019, \mn@doi [\aap] {10.1051/0004-6361/201935496},
  \href {https://ui.adsabs.harvard.edu/abs/2019A&A...626L...2F} {626, L2}

\bibitem[\protect\citeauthoryear{{Fang} et~al.,}{{Fang}
  et~al.}{2018}]{Fang2018}
{Fang} M.,  et~al., 2018, \mn@doi [\apj] {10.3847/1538-4357/aae780}, \href
  {https://ui.adsabs.harvard.edu/abs/2018ApJ...868...28F} {868, 28}

\bibitem[\protect\citeauthoryear{{Fedele}, {van den Ancker}, {Henning},
  {Jayawardhana}  \& {Oliveira}}{{Fedele} et~al.}{2010}]{Fedele2010}
{Fedele} D.,  {van den Ancker} M.~E.,  {Henning} T.,  {Jayawardhana} R.,
  {Oliveira} J.~M.,  2010, \mn@doi [\aap] {10.1051/0004-6361/200912810}, \href
  {https://ui.adsabs.harvard.edu/abs/2010A&A...510A..72F} {510, A72}

\bibitem[\protect\citeauthoryear{{Ferreira}}{{Ferreira}}{1997}]{Ferreira1997}
{Ferreira} J.,  1997, \aap, \href
  {https://ui.adsabs.harvard.edu/abs/1997A&A...319..340F} {319, 340}

\bibitem[\protect\citeauthoryear{{Ferreira} \& {Pelletier}}{{Ferreira} \&
  {Pelletier}}{1995}]{Ferreira&Pelletier1995}
{Ferreira} J.,  {Pelletier} G.,  1995, \aap, \href
  {https://ui.adsabs.harvard.edu/abs/1995A&A...295..807F} {295, 807}

\bibitem[\protect\citeauthoryear{{Flaherty} et~al.,}{{Flaherty}
  et~al.}{2017}]{Flaherty2017}
{Flaherty} K.~M.,  et~al., 2017, \mn@doi [\apj] {10.3847/1538-4357/aa79f9},
  \href {https://ui.adsabs.harvard.edu/abs/2017ApJ...843..150F} {843, 150}

\bibitem[\protect\citeauthoryear{{Flaherty}, {Hughes}, {Teague}, {Simon},
  {Andrews}  \& {Wilner}}{{Flaherty} et~al.}{2018}]{Flaherty2018}
{Flaherty} K.~M.,  {Hughes} A.~M.,  {Teague} R.,  {Simon} J.~B.,  {Andrews}
  S.~M.,   {Wilner} D.~J.,  2018, \mn@doi [\apj] {10.3847/1538-4357/aab615},
  \href {https://ui.adsabs.harvard.edu/abs/2018ApJ...856..117F} {856, 117}

\bibitem[\protect\citeauthoryear{{Flaherty} et~al.,}{{Flaherty}
  et~al.}{2020}]{Flaherty2020}
{Flaherty} K.,  et~al., 2020, \mn@doi [\apj] {10.3847/1538-4357/ab8cc5}, \href
  {https://ui.adsabs.harvard.edu/abs/2020ApJ...895..109F} {895, 109}

\bibitem[\protect\citeauthoryear{{Gaia Collaboration} et~al.,}{{Gaia
  Collaboration} et~al.}{2021}]{GAIAeDR3}
{Gaia Collaboration} et~al., 2021, \mn@doi [\aap]
  {10.1051/0004-6361/202039657}, \href
  {https://ui.adsabs.harvard.edu/abs/2021A&A...649A...1G} {649, A1}

\bibitem[\protect\citeauthoryear{{Gammie}}{{Gammie}}{1996}]{Gammie1996}
{Gammie} C.~F.,  1996, \mn@doi [\apj] {10.1086/176735}, \href
  {https://ui.adsabs.harvard.edu/abs/1996ApJ...457..355G} {457, 355}

\bibitem[\protect\citeauthoryear{{G{\'a}rate}, {Birnstiel},
  {Dr{\k{a}}{\.z}kowska}  \& {Stammler}}{{G{\'a}rate}
  et~al.}{2020}]{Garate2020}
{G{\'a}rate} M.,  {Birnstiel} T.,  {Dr{\k{a}}{\.z}kowska} J.,   {Stammler}
  S.~M.,  2020, \mn@doi [\aap] {10.1051/0004-6361/201936067}, \href
  {https://ui.adsabs.harvard.edu/abs/2020A&A...635A.149G} {635, A149}

\bibitem[\protect\citeauthoryear{{Giacalone}, {Teitler}, {K{\"o}nigl}, {Krijt}
  \& {Ciesla}}{{Giacalone} et~al.}{2019}]{Giacalone2019}
{Giacalone} S.,  {Teitler} S.,  {K{\"o}nigl} A.,  {Krijt} S.,   {Ciesla} F.~J.,
   2019, \mn@doi [\apj] {10.3847/1538-4357/ab311a}, \href
  {https://ui.adsabs.harvard.edu/abs/2019ApJ...882...33G} {882, 33}

\bibitem[\protect\citeauthoryear{{Gressel}, {Ramsey}, {Brinch}, {Nelson},
  {Turner}  \& {Bruderer}}{{Gressel} et~al.}{2020}]{Gressel2020}
{Gressel} O.,  {Ramsey} J.~P.,  {Brinch} C.,  {Nelson} R.~P.,  {Turner} N.~J.,
   {Bruderer} S.,  2020, \mn@doi [\apj] {10.3847/1538-4357/ab91b7}, \href
  {https://ui.adsabs.harvard.edu/abs/2020ApJ...896..126G} {896, 126}

\bibitem[\protect\citeauthoryear{{Gundlach} \& {Blum}}{{Gundlach} \&
  {Blum}}{2015}]{Gundlach&Blum2015}
{Gundlach} B.,  {Blum} J.,  2015, \mn@doi [\apj] {10.1088/0004-637X/798/1/34},
  \href {https://ui.adsabs.harvard.edu/abs/2015ApJ...798...34G} {798, 34}

\bibitem[\protect\citeauthoryear{{Harris} et~al.,}{{Harris}
  et~al.}{2020}]{numpy20_2020Natur.585..357H}
{Harris} C.~R.,  et~al., 2020, \mn@doi [\nat] {10.1038/s41586-020-2649-2},
  \href {https://ui.adsabs.harvard.edu/abs/2020Natur.585..357H} {585, 357}

\bibitem[\protect\citeauthoryear{{Harrison} et~al.,}{{Harrison}
  et~al.}{2021}]{Harrison2021}
{Harrison} R.~E.,  et~al., 2021, \mn@doi [\apj] {10.3847/1538-4357/abd94e},
  \href {https://ui.adsabs.harvard.edu/abs/2021ApJ...908..141H} {908, 141}

\bibitem[\protect\citeauthoryear{{Hartmann}, {Herczeg}  \& {Calvet}}{{Hartmann}
  et~al.}{2016}]{Hartmann2016}
{Hartmann} L.,  {Herczeg} G.,   {Calvet} N.,  2016, \mn@doi [\araa]
  {10.1146/annurev-astro-081915-023347}, \href
  {https://ui.adsabs.harvard.edu/abs/2016ARA&A..54..135H} {54, 135}

\bibitem[\protect\citeauthoryear{{Hendler}, {Pascucci}, {Pinilla}, {Tazzari},
  {Carpenter}, {Malhotra}  \& {Testi}}{{Hendler} et~al.}{2020}]{Hendler2020}
{Hendler} N.,  {Pascucci} I.,  {Pinilla} P.,  {Tazzari} M.,  {Carpenter} J.,
  {Malhotra} R.,   {Testi} L.,  2020, \mn@doi [\apj]
  {10.3847/1538-4357/ab70ba}, \href
  {https://ui.adsabs.harvard.edu/abs/2020ApJ...895..126H} {895, 126}

\bibitem[\protect\citeauthoryear{Hunter}{Hunter}{2007}]{matplotlib_Hunter:2007}
Hunter J.~D.,  2007, \mn@doi [Computing in Science \& Engineering]
  {10.1109/MCSE.2007.55}, 9, 90

\bibitem[\protect\citeauthoryear{{Kenyon} \& {Hartmann}}{{Kenyon} \&
  {Hartmann}}{1987}]{Kenyon&Hartmann1987}
{Kenyon} S.~J.,  {Hartmann} L.,  1987, \mn@doi [\apj] {10.1086/165866}, \href
  {https://ui.adsabs.harvard.edu/abs/1987ApJ...323..714K} {323, 714}

\bibitem[\protect\citeauthoryear{Kluyver et~al.,}{Kluyver
  et~al.}{2016}]{Jupyter_nootbok}
Kluyver T.,  et~al., 2016, in Loizides F.,  Scmidt B.,  eds, Positioning and
  Power in Academic Publishing: Players, Agents and Agendas. IOS Press, pp
  87--90, \url {https://eprints.soton.ac.uk/403913/}

\bibitem[\protect\citeauthoryear{{Koepferl}, {Ercolano}, {Dale}, {Teixeira},
  {Ratzka}  \& {Spezzi}}{{Koepferl} et~al.}{2013}]{Koepferl&Ercolano2013}
{Koepferl} C.~M.,  {Ercolano} B.,  {Dale} J.,  {Teixeira} P.~S.,  {Ratzka} T.,
   {Spezzi} L.,  2013, \mn@doi [\mnras] {10.1093/mnras/sts276}, \href
  {https://ui.adsabs.harvard.edu/abs/2013MNRAS.428.3327K} {428, 3327}

\bibitem[\protect\citeauthoryear{{Kratter} \& {Lodato}}{{Kratter} \&
  {Lodato}}{2016}]{Kratter&Lodato2016}
{Kratter} K.,  {Lodato} G.,  2016, \mn@doi [\araa]
  {10.1146/annurev-astro-081915-023307}, \href
  {https://ui.adsabs.harvard.edu/abs/2016ARA&A..54..271K} {54, 271}

\bibitem[\protect\citeauthoryear{{Laibe} \& {Price}}{{Laibe} \&
  {Price}}{2014}]{Laibe&Price2014}
{Laibe} G.,  {Price} D.~J.,  2014, \mn@doi [\mnras] {10.1093/mnras/stu1367},
  \href {https://ui.adsabs.harvard.edu/abs/2014MNRAS.444.1940L} {444, 1940}

\bibitem[\protect\citeauthoryear{{Lesur}}{{Lesur}}{2020}]{Lesur2020}
{Lesur} G.,  2020, arXiv e-prints, \href
  {https://ui.adsabs.harvard.edu/abs/2020arXiv200715967L} {p. arXiv:2007.15967}

\bibitem[\protect\citeauthoryear{{Lesur}}{{Lesur}}{2021}]{Lesur2021}
{Lesur} G. R.~J.,  2021, \mn@doi [\aap] {10.1051/0004-6361/202040109}, \href
  {https://ui.adsabs.harvard.edu/abs/2021A&A...650A..35L} {650, A35}

\bibitem[\protect\citeauthoryear{{Lodato}, {Scardoni}, {Manara}  \&
  {Testi}}{{Lodato} et~al.}{2017}]{Lodato2017}
{Lodato} G.,  {Scardoni} C.~E.,  {Manara} C.~F.,   {Testi} L.,  2017, \mn@doi
  [\mnras] {10.1093/mnras/stx2273}, \href
  {https://ui.adsabs.harvard.edu/abs/2017MNRAS.472.4700L} {472, 4700}

\bibitem[\protect\citeauthoryear{{Long} et~al.,}{{Long}
  et~al.}{2018a}]{Long2018_cham}
{Long} F.,  et~al., 2018a, \mn@doi [\apj] {10.3847/1538-4357/aacce9}, \href
  {https://ui.adsabs.harvard.edu/abs/2018ApJ...863...61L} {863, 61}

\bibitem[\protect\citeauthoryear{{Long} et~al.,}{{Long}
  et~al.}{2018b}]{Long2018}
{Long} F.,  et~al., 2018b, \mn@doi [\apj] {10.3847/1538-4357/aae8e1}, \href
  {https://ui.adsabs.harvard.edu/abs/2018ApJ...869...17L} {869, 17}

\bibitem[\protect\citeauthoryear{{Long} et~al.,}{{Long}
  et~al.}{2019}]{Long2019}
{Long} F.,  et~al., 2019, \mn@doi [\apj] {10.3847/1538-4357/ab2d2d}, \href
  {https://ui.adsabs.harvard.edu/abs/2019ApJ...882...49L} {882, 49}

\bibitem[\protect\citeauthoryear{{Louvet} et~al.,}{{Louvet}
  et~al.}{2016}]{Louvet2016}
{Louvet} F.,  et~al., 2016, \mn@doi [\aap] {10.1051/0004-6361/201628474}, \href
  {https://ui.adsabs.harvard.edu/abs/2016A&A...596A..88L} {596, A88}

\bibitem[\protect\citeauthoryear{{Louvet}, {Dougados}, {Cabrit}, {Mardones},
  {M{\'e}nard}, {Tabone}, {Pinte}  \& {Dent}}{{Louvet}
  et~al.}{2018}]{Louvet2018}
{Louvet} F.,  {Dougados} C.,  {Cabrit} S.,  {Mardones} D.,  {M{\'e}nard} F.,
  {Tabone} B.,  {Pinte} C.,   {Dent} W.~R.~F.,  2018, \mn@doi [\aap]
  {10.1051/0004-6361/201731733}, \href
  {https://ui.adsabs.harvard.edu/abs/2018A&A...618A.120L} {618, A120}

\bibitem[\protect\citeauthoryear{{Luhman} et~al.,}{{Luhman}
  et~al.}{2008}]{Luhman2008}
{Luhman} K.~L.,  et~al., 2008, \mn@doi [\apj] {10.1086/527347}, \href
  {https://ui.adsabs.harvard.edu/abs/2008ApJ...675.1375L} {675, 1375}

\bibitem[\protect\citeauthoryear{{Lynden-Bell} \& {Pringle}}{{Lynden-Bell} \&
  {Pringle}}{1974}]{Lynden-Bell&Pringle1974}
{Lynden-Bell} D.,  {Pringle} J.~E.,  1974, \mn@doi [\mnras]
  {10.1093/mnras/168.3.603}, \href
  {https://ui.adsabs.harvard.edu/abs/1974MNRAS.168..603L} {168, 603}

\bibitem[\protect\citeauthoryear{{Madhusudhan}}{{Madhusudhan}}{2019}]{Madhusudhan2019}
{Madhusudhan} N.,  2019, \mn@doi [\araa] {10.1146/annurev-astro-081817-051846},
  \href {https://ui.adsabs.harvard.edu/abs/2019ARA&A..57..617M} {57, 617}

\bibitem[\protect\citeauthoryear{{Manara} et~al.,}{{Manara}
  et~al.}{2016}]{Manara2016}
{Manara} C.~F.,  et~al., 2016, \mn@doi [\aap] {10.1051/0004-6361/201628549},
  \href {https://ui.adsabs.harvard.edu/abs/2016A&A...591L...3M} {591, L3}

\bibitem[\protect\citeauthoryear{{Manara} et~al.,}{{Manara}
  et~al.}{2019}]{Manara2019}
{Manara} C.~F.,  et~al., 2019, \mn@doi [\aap] {10.1051/0004-6361/201935964},
  \href {https://ui.adsabs.harvard.edu/abs/2019A&A...628A..95M} {628, A95}

\bibitem[\protect\citeauthoryear{{Manara}, {Ansdell}, {Rosotti}, {Hughes},
  {Armitage}, {Lodato}  \& {Williams}}{{Manara} et~al.}{2022}]{Manara2022}
{Manara} C.~F.,  {Ansdell} M.,  {Rosotti} G.~P.,  {Hughes} A.~M.,  {Armitage}
  P.~J.,  {Lodato} G.,   {Williams} J.~P.,  2022, arXiv e-prints, \href
  {https://ui.adsabs.harvard.edu/abs/2022arXiv220309930M} {p. arXiv:2203.09930}

\bibitem[\protect\citeauthoryear{{Maret} et~al.,}{{Maret}
  et~al.}{2020}]{Maret2020}
{Maret} S.,  et~al., 2020, \mn@doi [\aap] {10.1051/0004-6361/201936798}, \href
  {https://ui.adsabs.harvard.edu/abs/2020A&A...635A..15M} {635, A15}

\bibitem[\protect\citeauthoryear{{Mathis}, {Rumpl}  \& {Nordsieck}}{{Mathis}
  et~al.}{1977}]{Mathis1977}
{Mathis} J.~S.,  {Rumpl} W.,   {Nordsieck} K.~H.,  1977, \mn@doi [\apj]
  {10.1086/155591}, \href
  {https://ui.adsabs.harvard.edu/abs/1977ApJ...217..425M} {217, 425}

\bibitem[\protect\citeauthoryear{{Maury} et~al.,}{{Maury}
  et~al.}{2019}]{Maury2019}
{Maury} A.~J.,  et~al., 2019, \mn@doi [\aap] {10.1051/0004-6361/201833537},
  \href {https://ui.adsabs.harvard.edu/abs/2019A&A...621A..76M} {621, A76}

\bibitem[\protect\citeauthoryear{{Miotello}, {Testi}, {Lodato}, {Ricci},
  {Rosotti}, {Brooks}, {Maury}  \& {Natta}}{{Miotello}
  et~al.}{2014}]{Miotello2014}
{Miotello} A.,  {Testi} L.,  {Lodato} G.,  {Ricci} L.,  {Rosotti} G.,  {Brooks}
  K.,  {Maury} A.,   {Natta} A.,  2014, \mn@doi [\aap]
  {10.1051/0004-6361/201322945}, \href
  {https://ui.adsabs.harvard.edu/abs/2014A&A...567A..32M} {567, A32}

\bibitem[\protect\citeauthoryear{{Miotello}, {Rosotti}, {Ansdell}, {Facchini},
  {Manara}, {Williams}  \& {Bruderer}}{{Miotello} et~al.}{2021}]{Miotello2021}
{Miotello} A.,  {Rosotti} G.,  {Ansdell} M.,  {Facchini} S.,  {Manara} C.~F.,
  {Williams} J.~P.,   {Bruderer} S.,  2021, \mn@doi [\aap]
  {10.1051/0004-6361/202140550}, \href
  {https://ui.adsabs.harvard.edu/abs/2021A&A...651A..48M} {651, A48}

\bibitem[\protect\citeauthoryear{{Miotello}, {Kamp}, {Birnstiel}, {Cleeves}  \&
  {Kataoka}}{{Miotello} et~al.}{2022}]{Miotello2022}
{Miotello} A.,  {Kamp} I.,  {Birnstiel} T.,  {Cleeves} L.~I.,   {Kataoka} A.,
  2022, arXiv e-prints, \href
  {https://ui.adsabs.harvard.edu/abs/2022arXiv220309818M} {p. arXiv:2203.09818}

\bibitem[\protect\citeauthoryear{{Miyake}, {Suzuki}  \& {Inutsuka}}{{Miyake}
  et~al.}{2016}]{Miyake2016}
{Miyake} T.,  {Suzuki} T.~K.,   {Inutsuka} S.-i.,  2016, \mn@doi [\apj]
  {10.3847/0004-637X/821/1/3}, \href
  {https://ui.adsabs.harvard.edu/abs/2016ApJ...821....3M} {821, 3}

\bibitem[\protect\citeauthoryear{{Morbidelli} \& {Raymond}}{{Morbidelli} \&
  {Raymond}}{2016}]{Morbidelli&Raymond2016}
{Morbidelli} A.,  {Raymond} S.~N.,  2016, \mn@doi [Journal of Geophysical
  Research (Planets)] {10.1002/2016JE005088}, \href
  {https://ui.adsabs.harvard.edu/abs/2016JGRE..121.1962M} {121, 1962}

\bibitem[\protect\citeauthoryear{{Najita} \& {Bergin}}{{Najita} \&
  {Bergin}}{2018}]{Najita&Bergin2018}
{Najita} J.~R.,  {Bergin} E.~A.,  2018, \mn@doi [\apj]
  {10.3847/1538-4357/aad80c}, \href
  {https://ui.adsabs.harvard.edu/abs/2018ApJ...864..168N} {864, 168}

\bibitem[\protect\citeauthoryear{{{\"O}berg} \& {Bergin}}{{{\"O}berg} \&
  {Bergin}}{2021}]{Oberg&Bergin2021}
{{\"O}berg} K.~I.,  {Bergin} E.~A.,  2021, \mn@doi [\physrep]
  {10.1016/j.physrep.2020.09.004}, \href
  {https://ui.adsabs.harvard.edu/abs/2021PhR...893....1O} {893, 1}

\bibitem[\protect\citeauthoryear{{Owen}, {Ercolano}, {Clarke}  \&
  {Alexander}}{{Owen} et~al.}{2010}]{Owen2010}
{Owen} J.~E.,  {Ercolano} B.,  {Clarke} C.~J.,   {Alexander} R.~D.,  2010,
  \mn@doi [\mnras] {10.1111/j.1365-2966.2009.15771.x}, \href
  {https://ui.adsabs.harvard.edu/abs/2010MNRAS.401.1415O} {401, 1415}

\bibitem[\protect\citeauthoryear{{Pascucci} et~al.,}{{Pascucci}
  et~al.}{2016}]{Pascucci2016}
{Pascucci} I.,  et~al., 2016, \mn@doi [\apj] {10.3847/0004-637X/831/2/125},
  \href {https://ui.adsabs.harvard.edu/abs/2016ApJ...831..125P} {831, 125}

\bibitem[\protect\citeauthoryear{{Pascucci} et~al.,}{{Pascucci}
  et~al.}{2020}]{Pascucci2020}
{Pascucci} I.,  et~al., 2020, \mn@doi [\apj] {10.3847/1538-4357/abba3c}, \href
  {https://ui.adsabs.harvard.edu/abs/2020ApJ...903...78P} {903, 78}

\bibitem[\protect\citeauthoryear{{Pinilla}, {Birnstiel}, {Ricci}, {Dullemond},
  {Uribe}, {Testi}  \& {Natta}}{{Pinilla} et~al.}{2012}]{Pinilla12}
{Pinilla} P.,  {Birnstiel} T.,  {Ricci} L.,  {Dullemond} C.~P.,  {Uribe} A.~L.,
   {Testi} L.,   {Natta} A.,  2012, \mn@doi [\aap]
  {10.1051/0004-6361/201118204}, \href
  {https://ui.adsabs.harvard.edu/abs/2012A&A...538A.114P} {538, A114}

\bibitem[\protect\citeauthoryear{{Pinilla}, {Pascucci}  \& {Marino}}{{Pinilla}
  et~al.}{2020}]{Pinilla2020}
{Pinilla} P.,  {Pascucci} I.,   {Marino} S.,  2020, \mn@doi [\aap]
  {10.1051/0004-6361/201937003}, \href
  {https://ui.adsabs.harvard.edu/abs/2020A&A...635A.105P} {635, A105}

\bibitem[\protect\citeauthoryear{{Pinilla}, {Lenz}  \& {Stammler}}{{Pinilla}
  et~al.}{2021}]{Pinilla2021}
{Pinilla} P.,  {Lenz} C.~T.,   {Stammler} S.~M.,  2021, \mn@doi [\aap]
  {10.1051/0004-6361/202038920}, \href
  {https://ui.adsabs.harvard.edu/abs/2021A&A...645A..70P} {645, A70}

\bibitem[\protect\citeauthoryear{{Preibisch}, {Brown}, {Bridges}, {Guenther}
  \& {Zinnecker}}{{Preibisch} et~al.}{2002}]{Preibisch2002}
{Preibisch} T.,  {Brown} A. G.~A.,  {Bridges} T.,  {Guenther} E.,   {Zinnecker}
  H.,  2002, \mn@doi [\aj] {10.1086/341174}, \href
  {https://ui.adsabs.harvard.edu/abs/2002AJ....124..404P} {124, 404}

\bibitem[\protect\citeauthoryear{{Rodenkirch} \& {Dullemond}}{{Rodenkirch} \&
  {Dullemond}}{2022}]{Rodenkirch&Dullemond2022}
{Rodenkirch} P.~J.,  {Dullemond} C.~P.,  2022, \mn@doi [\aap]
  {10.1051/0004-6361/202142571}, \href
  {https://ui.adsabs.harvard.edu/abs/2022A&A...659A..42R} {659, A42}

\bibitem[\protect\citeauthoryear{{Rosotti}, {Clarke}, {Manara}  \&
  {Facchini}}{{Rosotti} et~al.}{2017}]{Rosotti2017}
{Rosotti} G.~P.,  {Clarke} C.~J.,  {Manara} C.~F.,   {Facchini} S.,  2017,
  \mn@doi [\mnras] {10.1093/mnras/stx595}, \href
  {https://ui.adsabs.harvard.edu/abs/2017MNRAS.468.1631R} {468, 1631}

\bibitem[\protect\citeauthoryear{{Rosotti}, {Booth}, {Tazzari}, {Clarke},
  {Lodato}  \& {Testi}}{{Rosotti} et~al.}{2019a}]{Rosotti2019_fr}
{Rosotti} G.~P.,  {Booth} R.~A.,  {Tazzari} M.,  {Clarke} C.,  {Lodato} G.,
  {Testi} L.,  2019a, \mn@doi [\mnras] {10.1093/mnrasl/slz064}, \href
  {https://ui.adsabs.harvard.edu/abs/2019MNRAS.486L..63R} {486, L63}

\bibitem[\protect\citeauthoryear{{Rosotti}, {Tazzari}, {Booth}, {Testi},
  {Lodato}  \& {Clarke}}{{Rosotti} et~al.}{2019b}]{Rosotti2019_radii}
{Rosotti} G.~P.,  {Tazzari} M.,  {Booth} R.~A.,  {Testi} L.,  {Lodato} G.,
  {Clarke} C.,  2019b, \mn@doi [\mnras] {10.1093/mnras/stz1190}, \href
  {https://ui.adsabs.harvard.edu/abs/2019MNRAS.486.4829R} {486, 4829}

\bibitem[\protect\citeauthoryear{{Safronov}}{{Safronov}}{1972}]{Safronov1969}
{Safronov} V.~S.,  1972, {Evolution of the protoplanetary cloud and formation
  of the earth and planets.}

\bibitem[\protect\citeauthoryear{{Salmeron} \& {Ireland}}{{Salmeron} \&
  {Ireland}}{2012}]{Salmeron&Ireland2012}
{Salmeron} R.,  {Ireland} T.~R.,  2012, \mn@doi [Earth and Planetary Science
  Letters] {10.1016/j.epsl.2012.01.033}, \href
  {https://ui.adsabs.harvard.edu/abs/2012E&PSL.327...61S} {327, 61}

\bibitem[\protect\citeauthoryear{{Sanchis} et~al.,}{{Sanchis}
  et~al.}{2021}]{Sanchis2021}
{Sanchis} E.,  et~al., 2021, \mn@doi [\aap] {10.1051/0004-6361/202039733},
  \href {https://ui.adsabs.harvard.edu/abs/2021A&A...649A..19S} {649, A19}

\bibitem[\protect\citeauthoryear{{Sellek}, {Booth}  \& {Clarke}}{{Sellek}
  et~al.}{2020a}]{Sellek2020}
{Sellek} A.~D.,  {Booth} R.~A.,   {Clarke} C.~J.,  2020a, \mn@doi [\mnras]
  {10.1093/mnras/stz3528}, \href
  {https://ui.adsabs.harvard.edu/abs/2020MNRAS.492.1279S} {492, 1279}

\bibitem[\protect\citeauthoryear{{Sellek}, {Booth}  \& {Clarke}}{{Sellek}
  et~al.}{2020b}]{Sellek2020_acc}
{Sellek} A.~D.,  {Booth} R.~A.,   {Clarke} C.~J.,  2020b, \mn@doi [\mnras]
  {10.1093/mnras/staa2519}, \href
  {https://ui.adsabs.harvard.edu/abs/2020MNRAS.498.2845S} {498, 2845}

\bibitem[\protect\citeauthoryear{{Shakura} \& {Sunyaev}}{{Shakura} \&
  {Sunyaev}}{1973}]{Shakura&Sunyaev1973}
{Shakura} N.~I.,  {Sunyaev} R.~A.,  1973, \aap, \href
  {https://ui.adsabs.harvard.edu/abs/1973A&A....24..337S} {500, 33}

\bibitem[\protect\citeauthoryear{{Suzuki}, {Ogihara}, {Morbidelli}, {Crida}  \&
  {Guillot}}{{Suzuki} et~al.}{2016}]{Suzuki2016}
{Suzuki} T.~K.,  {Ogihara} M.,  {Morbidelli} A.,  {Crida} A.,   {Guillot} T.,
  2016, \mn@doi [\aap] {10.1051/0004-6361/201628955}, \href
  {https://ui.adsabs.harvard.edu/abs/2016A&A...596A..74S} {596, A74}

\bibitem[\protect\citeauthoryear{{Tabone} et~al.,}{{Tabone}
  et~al.}{2017}]{Tabone2017}
{Tabone} B.,  et~al., 2017, \mn@doi [\aap] {10.1051/0004-6361/201731691}, \href
  {https://ui.adsabs.harvard.edu/abs/2017A&A...607L...6T} {607, L6}

\bibitem[\protect\citeauthoryear{{Tabone}, {Rosotti}, {Lodato}, {Armitage},
  {Cridland}  \& {van Dishoeck}}{{Tabone} et~al.}{2022a}]{Tabone2022b}
{Tabone} B.,  {Rosotti} G.~P.,  {Lodato} G.,  {Armitage} P.~J.,  {Cridland}
  A.~J.,   {van Dishoeck} E.~F.,  2022a, \mn@doi [\mnras]
  {10.1093/mnrasl/slab124}, \href
  {https://ui.adsabs.harvard.edu/abs/2022MNRAS.512L..74T} {512, L74}

\bibitem[\protect\citeauthoryear{{Tabone}, {Rosotti}, {Cridland}, {Armitage}
  \& {Lodato}}{{Tabone} et~al.}{2022b}]{Tabone2022a}
{Tabone} B.,  {Rosotti} G.~P.,  {Cridland} A.~J.,  {Armitage} P.~J.,   {Lodato}
  G.,  2022b, \mn@doi [\mnras] {10.1093/mnras/stab3442}, \href
  {https://ui.adsabs.harvard.edu/abs/2022MNRAS.512.2290T} {512, 2290}

\bibitem[\protect\citeauthoryear{{Takahashi} \& {Muto}}{{Takahashi} \&
  {Muto}}{2018}]{Takahashi&Muto2018}
{Takahashi} S.~Z.,  {Muto} T.,  2018, \mn@doi [\apj]
  {10.3847/1538-4357/aadda0}, \href
  {https://ui.adsabs.harvard.edu/abs/2018ApJ...865..102T} {865, 102}

\bibitem[\protect\citeauthoryear{{Takeuchi} \& {Lin}}{{Takeuchi} \&
  {Lin}}{2002}]{Takeuchi&Lin2002}
{Takeuchi} T.,  {Lin} D.~N.~C.,  2002, \mn@doi [\apj] {10.1086/344437}, \href
  {https://ui.adsabs.harvard.edu/abs/2002ApJ...581.1344T} {581, 1344}

\bibitem[\protect\citeauthoryear{{Taki}, {Kuwabara}, {Kobayashi}  \&
  {Suzuki}}{{Taki} et~al.}{2021}]{Taki2021}
{Taki} T.,  {Kuwabara} K.,  {Kobayashi} H.,   {Suzuki} T.~K.,  2021, \mn@doi
  [\apj] {10.3847/1538-4357/abd79f}, \href
  {https://ui.adsabs.harvard.edu/abs/2021ApJ...909...75T} {909, 75}

\bibitem[\protect\citeauthoryear{{Tanaka}, {Himeno}  \& {Ida}}{{Tanaka}
  et~al.}{2005}]{Tanaka2005}
{Tanaka} H.,  {Himeno} Y.,   {Ida} S.,  2005, \mn@doi [\apj] {10.1086/429658},
  \href {https://ui.adsabs.harvard.edu/abs/2005ApJ...625..414T} {625, 414}

\bibitem[\protect\citeauthoryear{{Tazzari} et~al.,}{{Tazzari}
  et~al.}{2016}]{Tazzari2016}
{Tazzari} M.,  et~al., 2016, \mn@doi [\aap] {10.1051/0004-6361/201527423},
  \href {https://ui.adsabs.harvard.edu/abs/2016A&A...588A..53T} {588, A53}

\bibitem[\protect\citeauthoryear{{Tazzari} et~al.,}{{Tazzari}
  et~al.}{2017}]{Tazzari2017}
{Tazzari} M.,  et~al., 2017, \mn@doi [\aap] {10.1051/0004-6361/201730890},
  \href {https://ui.adsabs.harvard.edu/abs/2017A&A...606A..88T} {606, A88}

\bibitem[\protect\citeauthoryear{{Tazzari}, {Clarke}, {Testi}, {Williams},
  {Facchini}, {Manara}, {Natta}  \& {Rosotti}}{{Tazzari}
  et~al.}{2021a}]{Tazzari2021_models}
{Tazzari} M.,  {Clarke} C.~J.,  {Testi} L.,  {Williams} J.~P.,  {Facchini} S.,
  {Manara} C.~F.,  {Natta} A.,   {Rosotti} G.,  2021a, \mn@doi [\mnras]
  {10.1093/mnras/stab1808}, \href
  {https://ui.adsabs.harvard.edu/abs/2021MNRAS.506.2804T} {506, 2804}

\bibitem[\protect\citeauthoryear{{Tazzari} et~al.,}{{Tazzari}
  et~al.}{2021b}]{Tazzari2021_obsv}
{Tazzari} M.,  et~al., 2021b, \mn@doi [\mnras] {10.1093/mnras/stab1912}, \href
  {https://ui.adsabs.harvard.edu/abs/2021MNRAS.506.5117T} {506, 5117}

\bibitem[\protect\citeauthoryear{{Testi} et~al.,}{{Testi}
  et~al.}{2014}]{Testi2014}
{Testi} L.,  et~al., 2014, in {Beuther} H.,  {Klessen} R.~S.,  {Dullemond}
  C.~P.,   {Henning} T.,  eds, Protostars and Planets VI. p.~339 (\mn@eprint
  {arXiv} {1402.1354}), \mn@doi{10.2458/azu\_uapress\_9780816531240-ch015}

\bibitem[\protect\citeauthoryear{{Tobin} et~al.,}{{Tobin}
  et~al.}{2020}]{Tobin2020}
{Tobin} J.~J.,  et~al., 2020, \mn@doi [\apj] {10.3847/1538-4357/ab6f64}, \href
  {https://ui.adsabs.harvard.edu/abs/2020ApJ...890..130T} {890, 130}

\bibitem[\protect\citeauthoryear{{Toci}, {Rosotti}, {Lodato}, {Testi}  \&
  {Trapman}}{{Toci} et~al.}{2021}]{Toci2021}
{Toci} C.,  {Rosotti} G.,  {Lodato} G.,  {Testi} L.,   {Trapman} L.,  2021,
  \mn@doi [\mnras] {10.1093/mnras/stab2112}, \href
  {https://ui.adsabs.harvard.edu/abs/2021MNRAS.507..818T} {507, 818}

\bibitem[\protect\citeauthoryear{{Trapman}, {Facchini}, {Hogerheijde}, {van
  Dishoeck}  \& {Bruderer}}{{Trapman} et~al.}{2019}]{Trapman2019}
{Trapman} L.,  {Facchini} S.,  {Hogerheijde} M.~R.,  {van Dishoeck} E.~F.,
  {Bruderer} S.,  2019, \mn@doi [\aap] {10.1051/0004-6361/201834723}, \href
  {https://ui.adsabs.harvard.edu/abs/2019A&A...629A..79T} {629, A79}

\bibitem[\protect\citeauthoryear{{Trapman}, {Rosotti}, {Bosman}, {Hogerheijde}
  \& {van Dishoeck}}{{Trapman} et~al.}{2020}]{Trapman2020}
{Trapman} L.,  {Rosotti} G.,  {Bosman} A.~D.,  {Hogerheijde} M.~R.,   {van
  Dishoeck} E.~F.,  2020, \mn@doi [\aap] {10.1051/0004-6361/202037673}, \href
  {https://ui.adsabs.harvard.edu/abs/2020A&A...640A...5T} {640, A5}

\bibitem[\protect\citeauthoryear{{Trapman}, {Tabone}, {Rosotti}  \&
  {Zhang}}{{Trapman} et~al.}{2022}]{Trapman2022}
{Trapman} L.,  {Tabone} B.,  {Rosotti} G.,   {Zhang} K.,  2022, \mn@doi [\apj]
  {10.3847/1538-4357/ac3ed5}, \href
  {https://ui.adsabs.harvard.edu/abs/2022ApJ...926...61T} {926, 61}

\bibitem[\protect\citeauthoryear{{Tripathi}, {Andrews}, {Birnstiel}  \&
  {Wilner}}{{Tripathi} et~al.}{2017}]{Tripathi2017}
{Tripathi} A.,  {Andrews} S.~M.,  {Birnstiel} T.,   {Wilner} D.~J.,  2017,
  \mn@doi [\apj] {10.3847/1538-4357/aa7c62}, \href
  {https://ui.adsabs.harvard.edu/abs/2017ApJ...845...44T} {845, 44}

\bibitem[\protect\citeauthoryear{{Turner}, {Fromang}, {Gammie}, {Klahr},
  {Lesur}, {Wardle}  \& {Bai}}{{Turner} et~al.}{2014}]{Turner2014}
{Turner} N.~J.,  {Fromang} S.,  {Gammie} C.,  {Klahr} H.,  {Lesur} G.,
  {Wardle} M.,   {Bai} X.~N.,  2014, in {Beuther} H.,  {Klessen} R.~S.,
  {Dullemond} C.~P.,   {Henning} T.,  eds, Protostars and Planets VI. p.~411
  (\mn@eprint {arXiv} {1401.7306}),
  \mn@doi{10.2458/azu\_uapress\_9780816531240-ch018}

\bibitem[\protect\citeauthoryear{{Virtanen} et~al.,}{{Virtanen}
  et~al.}{2020}]{scipy_2020SciPy-NMeth}
{Virtanen} P.,  et~al., 2020, \mn@doi [Nature Methods]
  {https://doi.org/10.1038/s41592-019-0686-2}, \href {https://rdcu.be/b08Wh}
  {17, 261}

\bibitem[\protect\citeauthoryear{{Vlemmings} et~al.,}{{Vlemmings}
  et~al.}{2019}]{Vlemmings2019}
{Vlemmings} W.~H.~T.,  et~al., 2019, \mn@doi [\aap]
  {10.1051/0004-6361/201935459}, \href
  {https://ui.adsabs.harvard.edu/abs/2019A&A...624L...7V} {624, L7}

\bibitem[\protect\citeauthoryear{{Wang}, {Bai}  \& {Goodman}}{{Wang}
  et~al.}{2019}]{Wang2019}
{Wang} L.,  {Bai} X.-N.,   {Goodman} J.,  2019, \mn@doi [\apj]
  {10.3847/1538-4357/ab06fd}, \href
  {https://ui.adsabs.harvard.edu/abs/2019ApJ...874...90W} {874, 90}

\bibitem[\protect\citeauthoryear{{Wardle} \& {Koenigl}}{{Wardle} \&
  {Koenigl}}{1993}]{Wardle&Koenigl1993}
{Wardle} M.,  {Koenigl} A.,  1993, \mn@doi [\apj] {10.1086/172739}, \href
  {https://ui.adsabs.harvard.edu/abs/1993ApJ...410..218W} {410, 218}

\bibitem[\protect\citeauthoryear{Waskom}{Waskom}{2021}]{Waskom2021}
Waskom M.~L.,  2021, \mn@doi [Journal of Open Source Software]
  {10.21105/joss.03021}, 6, 3021

\bibitem[\protect\citeauthoryear{{Weidenschilling}}{{Weidenschilling}}{1977}]{Weidenchilling1977}
{Weidenschilling} S.~J.,  1977, \mn@doi [\mnras] {10.1093/mnras/180.2.57},
  \href {https://ui.adsabs.harvard.edu/abs/1977MNRAS.180...57W} {180, 57}

\bibitem[\protect\citeauthoryear{{Whelan}, {Pascucci}, {Gorti}, {Edwards},
  {Alexander}, {Sterzik}  \& {Melo}}{{Whelan} et~al.}{2021}]{Whelan2021}
{Whelan} E.~T.,  {Pascucci} I.,  {Gorti} U.,  {Edwards} S.,  {Alexander} R.~D.,
   {Sterzik} M.~F.,   {Melo} C.,  2021, \mn@doi [\apj]
  {10.3847/1538-4357/abf55e}, \href
  {https://ui.adsabs.harvard.edu/abs/2021ApJ...913...43W} {913, 43}

\bibitem[\protect\citeauthoryear{{Whipple}}{{Whipple}}{1972}]{Whipple1972}
{Whipple} F.~L.,  1972, in {Elvius} A.,  ed., From Plasma to Planet. p.~211

\bibitem[\protect\citeauthoryear{{Winn} \& {Fabrycky}}{{Winn} \&
  {Fabrycky}}{2015}]{Winn&Fabricky2015}
{Winn} J.~N.,  {Fabrycky} D.~C.,  2015, \mn@doi [\araa]
  {10.1146/annurev-astro-082214-122246}, \href
  {https://ui.adsabs.harvard.edu/abs/2015ARA&A..53..409W} {53, 409}

\bibitem[\protect\citeauthoryear{{Wong}, {Hirashita}  \& {Li}}{{Wong}
  et~al.}{2016}]{Wong2016}
{Wong} Y. H.~V.,  {Hirashita} H.,   {Li} Z.-Y.,  2016, \mn@doi [\pasj]
  {10.1093/pasj/psw066}, \href
  {https://ui.adsabs.harvard.edu/abs/2016PASJ...68...67W} {68, 67}

\bibitem[\protect\citeauthoryear{{Youdin} \& {Lithwick}}{{Youdin} \&
  {Lithwick}}{2007}]{Youdin&Lithwick2007}
{Youdin} A.~N.,  {Lithwick} Y.,  2007, \mn@doi [\icarus]
  {10.1016/j.icarus.2007.07.012}, \href
  {https://ui.adsabs.harvard.edu/abs/2007Icar..192..588Y} {192, 588}

\bibitem[\protect\citeauthoryear{{Zagaria}, {Rosotti}  \& {Lodato}}{{Zagaria}
  et~al.}{2021}]{Zagaria2021_theory}
{Zagaria} F.,  {Rosotti} G.~P.,   {Lodato} G.,  2021, \mn@doi [\mnras]
  {10.1093/mnras/stab985}, \href
  {https://ui.adsabs.harvard.edu/abs/2021MNRAS.504.2235Z} {504, 2235}

\bibitem[\protect\citeauthoryear{{Zhu} et~al.,}{{Zhu} et~al.}{2019}]{Zhu2019}
{Zhu} Z.,  et~al., 2019, \mn@doi [\apjl] {10.3847/2041-8213/ab1f8c}, \href
  {https://ui.adsabs.harvard.edu/abs/2019ApJ...877L..18Z} {877, L18}

\bibitem[\protect\citeauthoryear{{Zormpas}, {Birnstiel}, {Rosotti}  \&
  {Andrews}}{{Zormpas} et~al.}{2022}]{Zormpas2022}
{Zormpas} A.,  {Birnstiel} T.,  {Rosotti} G.~P.,   {Andrews} S.~M.,  2022,
  arXiv e-prints, \href {https://ui.adsabs.harvard.edu/abs/2022arXiv220201241Z}
  {p. arXiv:2202.01241}

\bibitem[\protect\citeauthoryear{{de Valon}, {Dougados}, {Cabrit}, {Louvet},
  {Zapata}  \& {Mardones}}{{de Valon} et~al.}{2020}]{deValon2020}
{de Valon} A.,  {Dougados} C.,  {Cabrit} S.,  {Louvet} F.,  {Zapata} L.~A.,
  {Mardones} D.,  2020, \mn@doi [\aap] {10.1051/0004-6361/201936950}, \href
  {https://ui.adsabs.harvard.edu/abs/2020A&A...634L..12D} {634, L12}

\bibitem[\protect\citeauthoryear{{van der Marel} et~al.,}{{van der Marel}
  et~al.}{2018}]{vanderMarel2018}
{van der Marel} N.,  et~al., 2018, \mn@doi [\apj] {10.3847/1538-4357/aaaa6b},
  \href {https://ui.adsabs.harvard.edu/abs/2018ApJ...854..177V} {854, 177}

\makeatother
\end{thebibliography}



\appendix

\section{Code convergence test}\label{app:1}
\begin{figure*}
    \centering
    \includegraphics[width=\textwidth]{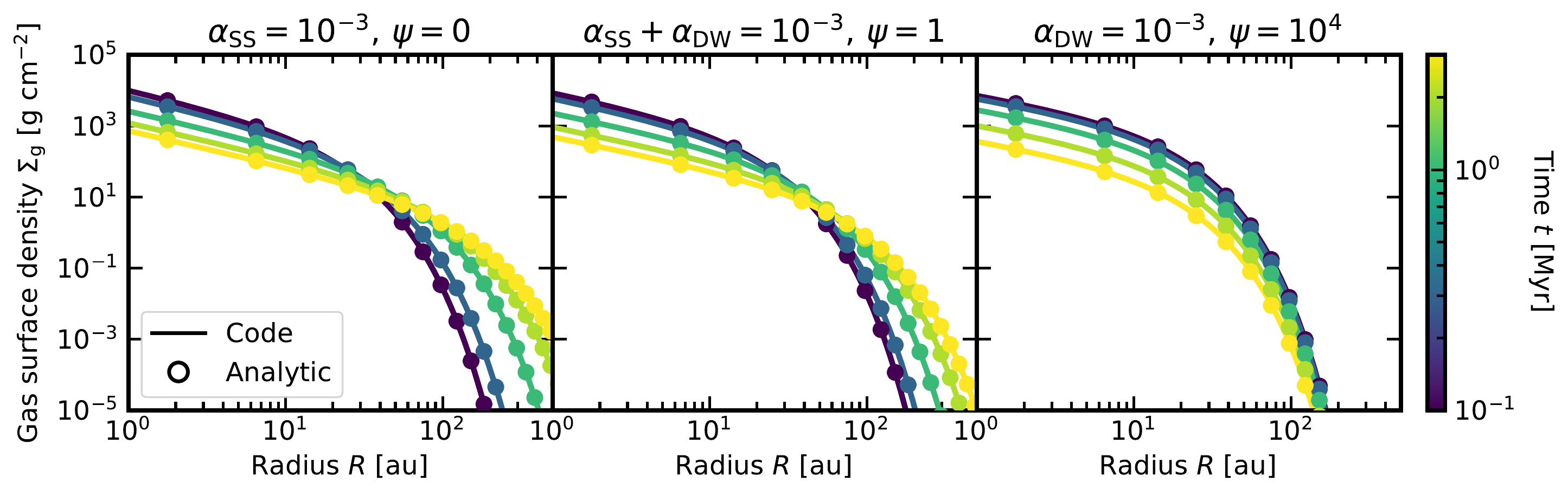}
    \caption{Code convergence test: agreement between the numerical solution and the analytical results of \citet{Tabone2022a}.}
    \label{fig:App1}
\end{figure*}

Fig.~\ref{fig:App1} displays the results of the code convergence test in the case of viscous-only (left panel), hybrid (middle panel) and purely MHD wind (right panel) models. At each time step an excellent agreement can be seen between the numerical and the analytical solutions, plotted as solid lines and dots, respectively.

\section{Dependence on the lever arm parameter}\label{app:3}
\begin{figure*}
    \centering
    \includegraphics[width=\textwidth]{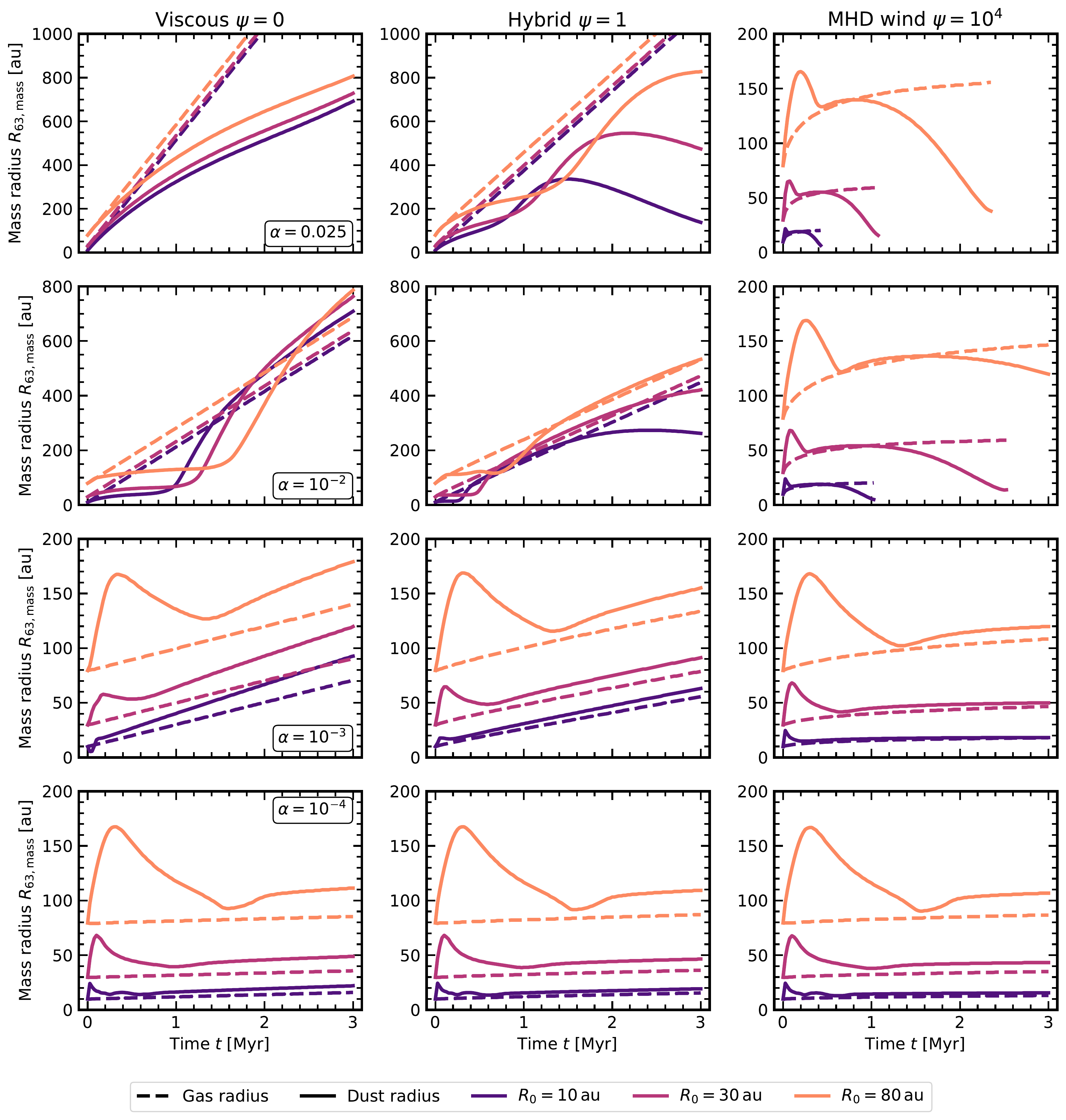}
    \caption{Mass radius for gas (dashed lines) and dust (solid lines) time evolution as a function of $\alpha$ (row by row) and $\psi$ (column by column), for $\lambda=1.5$.}
    \label{fig:App3.1}
\end{figure*}

\paragraph*{Mass radius (see Fig.~\ref{fig:App3.1})} Similar considerations as in Section~\ref{sec:3} apply. The gas radius is not constant in purely magnetic wind models because our initial condition is not the same as in \citet{Tabone2022a} and the discs undergo a phase of re-adjustment to match the radial profile typical of discs evolving under the effect of magnetic winds. Despite this difference, we decided to use Eq.~\ref{eq:2} regardless of $\psi$ and $\lambda$ to evolve all the models from the same initial conditions. Despite $\Sigma_{\rm d}$ being steeper than $\Sigma_{\rm g}$ in the wind-dominated models ($\xi=1$), still the gas radius is smaller than the dust radius because of the exponential tail. Using Eq.~39 in \citet{Birnstiel2012}, $\Sigma_{\rm g}\propto\exp{(-R/R_0)}$ and $\Sigma_{\rm d}\propto R^{-1/4}\exp{(-R/2R_0)}$, suggesting that the e-folding length of dust decay in the outer disc is larger than the gas one.

\begin{figure*}
    \centering
    \includegraphics[width=\textwidth]{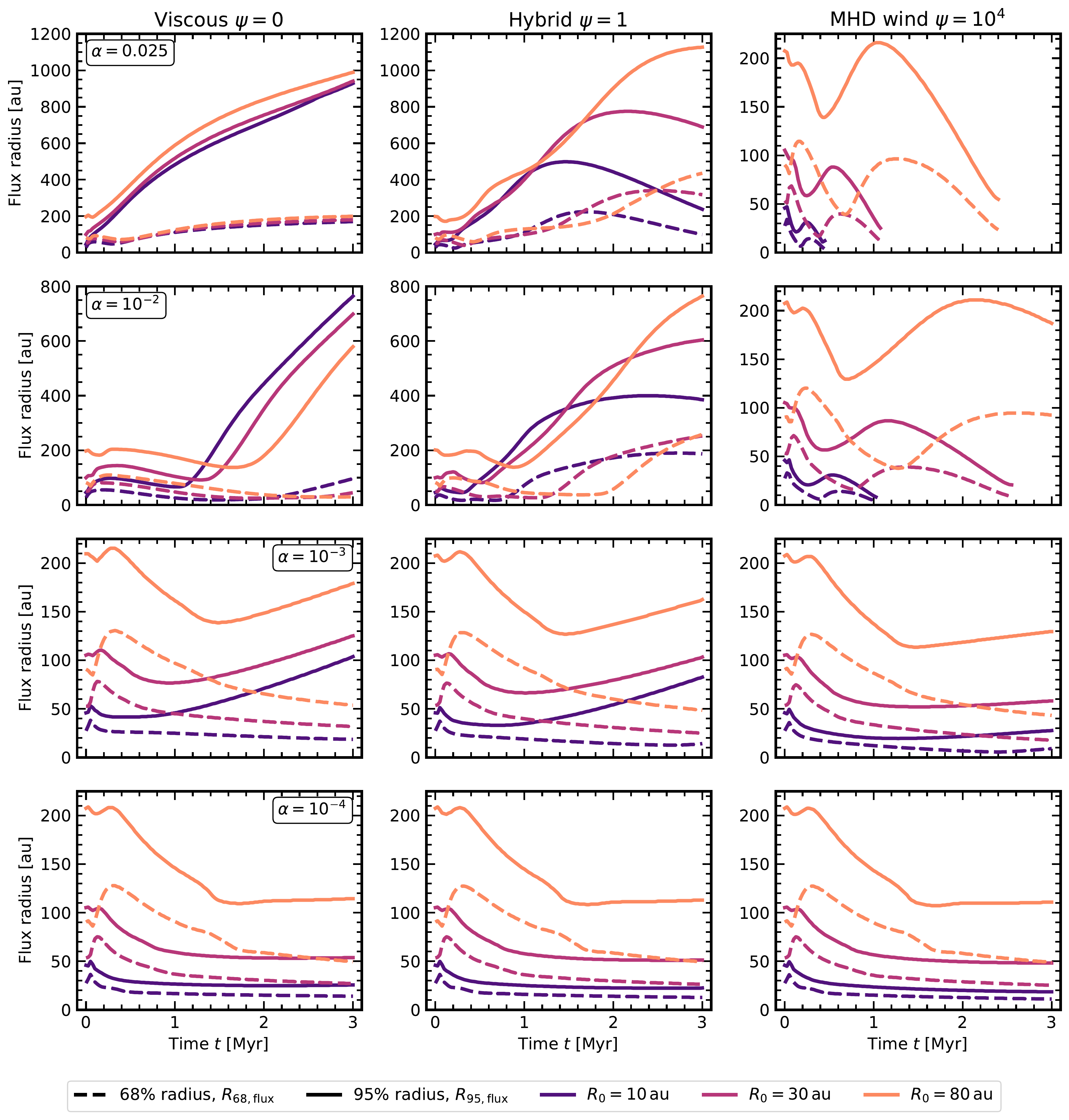}
    \caption{68 per cent (dashed lines) and 95 per cent (solid lines) flux radius time evolution as a function of $\alpha$ (row by row) and $\psi$ (column by column), for $\lambda=1.5$.}
    \label{fig:App3.2}
\end{figure*}

\begin{figure*}
    \centering
    \includegraphics[width=\textwidth]{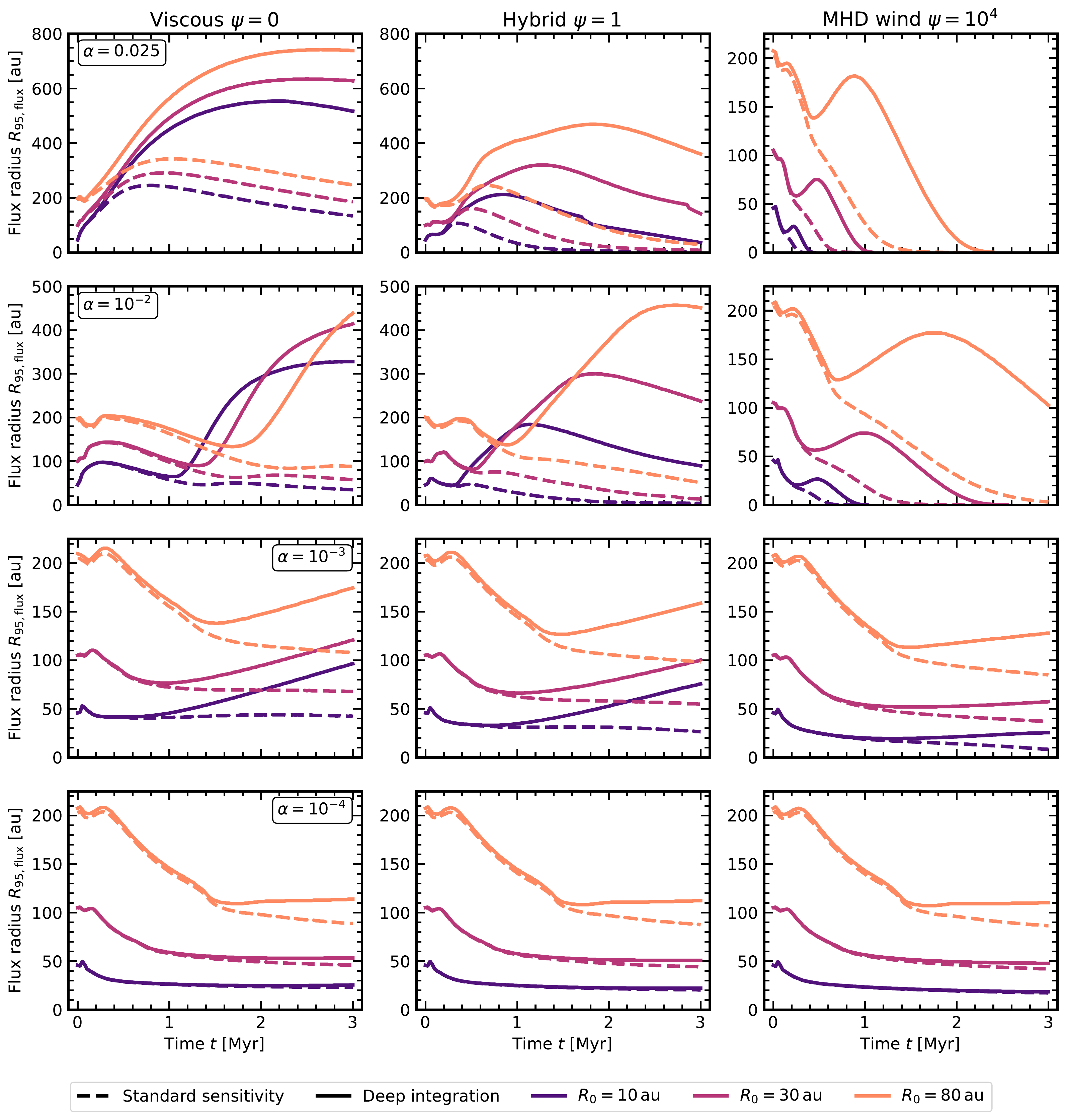}
    \caption{95 per cent flux radius time evolution as a function of $\alpha$ (row by row) and $\psi$ (column by column), for $\lambda=1.5$. The current survey case and the deep sensitivity scenario are plotted as dashed and solid lines.}
    \label{fig:App3.3}
\end{figure*}

\paragraph*{Flux radius (see Fig.s~\ref{fig:App3.2} and~\ref{fig:App3.3})} The main differences with Fig.~\ref{fig:4.2} are visible after $t\approx2\,{\rm Myr}$ when the faster removal of grains and less steep surface brightness promoted by $\lambda=1.5$, allows for larger disc sizes in the hybrid models with $\alpha\geq0.01$. On the contrary, when a sensitivity cut is included, those discs become much smaller than in Fig.~\ref{fig:5.1}. In the case of purely magnetic wind models with $\alpha=10^{-3}$ a late-time increase of the disc sizes can be observed, at odds with Fig.s~\ref{fig:4.2} and~\ref{fig:5.1}. Nevertheless, viscous and purely magnetic wind models can be still distinguished on a longer time scale, because in the latter case dust disc sizes stall by $10\,{\rm Myr}$.

\section{Finite dispersal time scale}\label{app:2}
\begin{figure*}
    \centering
    \includegraphics[width=\textwidth]{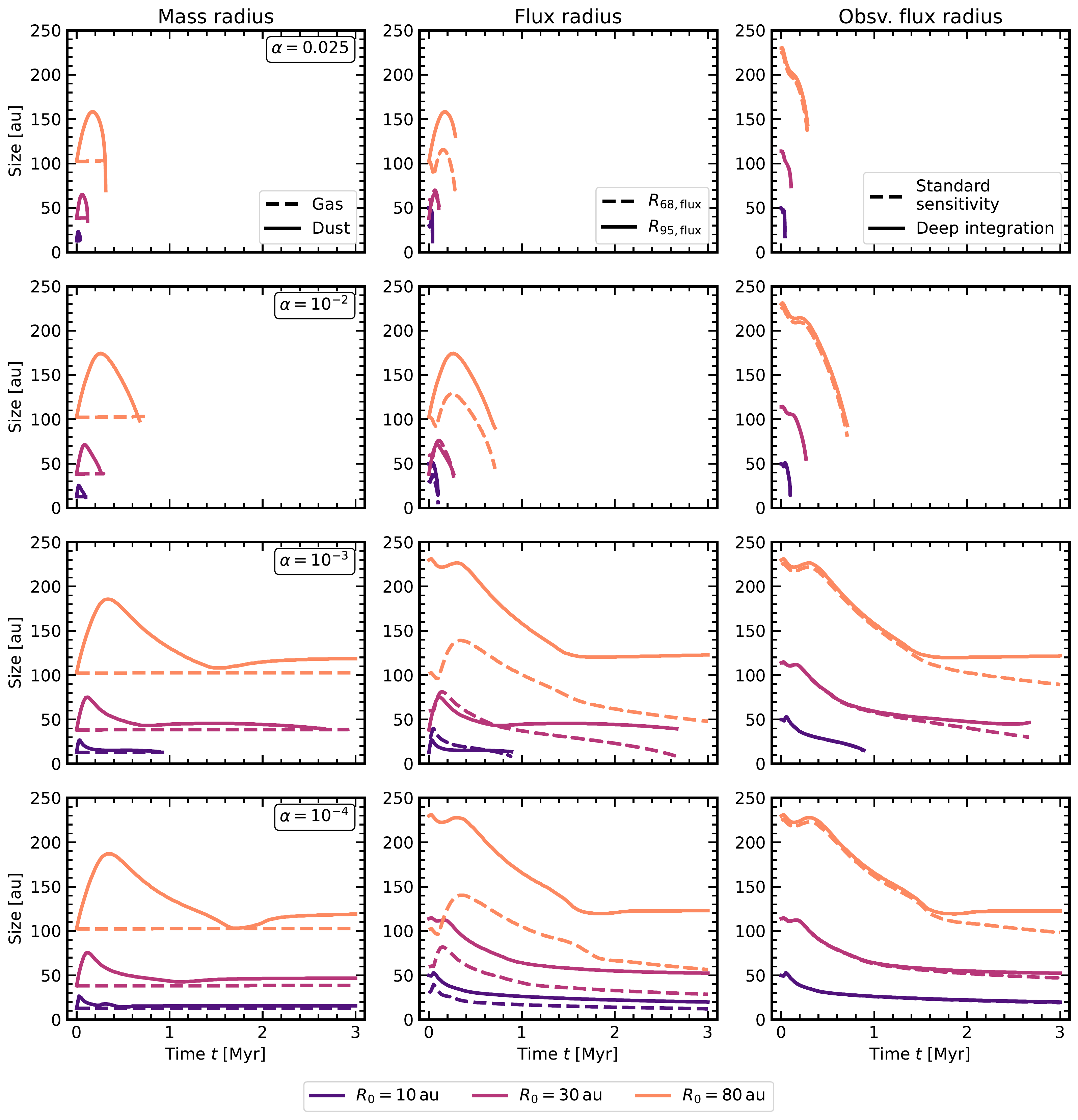}
    \caption{Disc size evolution assuming $\omega=1$ and $\lambda=3$.}
    \label{fig:App2}
\end{figure*}

In this Section we discuss the results of a set of purely magnetic wind models where $\alpha_{\rm DW}=\alpha_{\rm DW}(t=0)(M_{\rm disc}/M_0)^{-\omega}$ increases with time \citep{Tabone2022a}. In these models, when only gas is considered, the main disc properties, such as masses, accretion rates and sizes, do not change much until a significant fraction of the dispersal time scale, $t_{\rm disp}=2t_{\rm acc}/\omega$, elapsed. Instead, when solids are included, the corresponding \textit{dust} properties undergo relatively large variations. Fig.~\ref{fig:App2} shows it in the case of the mass and flux sizes, plotted for $\omega=1$ and different $\alpha_{\rm DW}(t=0)$ and initial sizes. Clearly, the growth and drift of solids largely impacts the dust disc sizes and before dispersal such sizes are similar to those in the $\omega=0$ case studied in the main text. 

What changes with higher values of $\omega$ is how fast the disc disperses. For this reason, a proper comparison among models with different values of $\lambda$ requires using the initial conditions of \citet{Tabone2022a} as in Fig.~\ref{fig:App2}. In fact, in this case $\lambda$ sets only the ratio of the mass loss rate in the wind to the mass accretion rate and does not influence the disc gas mass and dispersal time scale. As in the cases discussed in Appendix~\ref{app:3}, when $\lambda=1.5$ the disc dust sizes are larger than for $\lambda=3$, but follow the same evolutionary trend.

\bsp	
\label{lastpage}
\end{document}